\documentclass[11pt]{article}
\usepackage{graphicx}
\usepackage{amsmath}
\usepackage{amssymb}
\usepackage{caption2}
\begin{document}
\bibliographystyle{prsty}
\begin{center}
{\large {\bf \sc{ Possible pentaquark candidates: new excited $\Omega_c$ states }}} \\[2mm]
Zhi-Gang  Wang\footnote{E-mail: zgwang@aliyun.com; zgwang@ncepu.edu.cn. }, Jun-Xia Zhang    \\
 Department of Physics, North China Electric Power University,
Baoding 071003, P. R. China
\end{center}

\begin{abstract}
In this article, we  study  the axialvector-diquark-scalar-diquark-antiquark type charmed pentaquark states with $J^P={\frac{3}{2}}^\pm$
with the QCD sum rules by carrying out the operator product expansion   up to   the vacuum condensates of dimension $13$ in a consistent way.
In calculations,   we separate  the contributions of the negative parity and positive parity pentaquark states  unambiguously,
and choose three sets input parameters to study the masses and pole residues of the charmed pentaquark states $uuuc\bar{u}$ and $sssc\bar{s}$
  in details.  Then we estimate the masses of the charmed pentaquark states $ssuc\bar{u}$, $susc\bar{u}$, $ssdc\bar{d}$ and $sdsc\bar{d}$
  with $J^P={\frac{3}{2}}^-$ to be $3.15\pm0.13\,\rm{GeV}$ according to the $SU(3)$ breaking effects, which is compatible with
   the $\Omega_c(3050)$, $\Omega_c(3066)$, $\Omega_c(3090)$,
  $\Omega_c(3119)$.
\end{abstract}

 PACS number: 14.20.Lq, 14.20.Pt

 Key words: Charmed  baryon states, Charmed pentaquark states,  QCD sum rules

\section{Introduction}

In 2017, the LHCb collaboration studied the  $\Xi_c^+ K^-$  mass spectrum with a sample of $pp$
collision data corresponding to an integrated luminosity of  $3.3\rm{fb}^{-1}$   collected by the LHCb experiment
and observed five new narrow excited $\Omega_c$ states,
$\Omega_c(3000)$, $\Omega_c(3050)$, $\Omega_c(3066)$, $\Omega_c(3090)$, $\Omega_c(3119)$ \cite{LHCb-Omega}. The measured masses and widths are
\begin{flalign}
 & \Omega_c(3000) : M = 3000.4 \pm 0.2 \pm 0.1 \mbox{ MeV}\, , \, \Gamma = 4.5\pm0.6\pm0.3 \mbox{ MeV} \, , \nonumber \\
 & \Omega_c(3050) : M = 3050.2 \pm 0.1 \pm 0.1 \mbox{ MeV}\, , \, \Gamma = 0.8\pm0.2\pm0.1 \mbox{ MeV} \, , \nonumber \\
 & \Omega_c(3066) : M = 3065.6 \pm 0.1 \pm 0.3 \mbox{ MeV}\, , \, \Gamma = 3.5\pm0.4\pm0.2 \mbox{ MeV} \, , \nonumber \\
 & \Omega_c(3090) : M = 3090.2 \pm 0.3 \pm 0.5 \mbox{ MeV}\, , \, \Gamma = 8.7\pm1.0\pm0.8 \mbox{ MeV} \, , \nonumber \\
 & \Omega_c(3119) : M = 3119.1 \pm 0.3 \pm 0.9 \mbox{ MeV}\, , \, \Gamma = 1.1\pm0.8\pm0.4 \mbox{ MeV} \, .
\end{flalign}
Later, the Belle collaboration confirmed the excited $\Omega_c$ states  $\Omega_c(3000)$, $\Omega_c(3050)$, $\Omega_c(3066)$ and $\Omega_c(3090)$ in the decay mode $\Xi_c^+K^-$
 using the entire Belle data sample of $980\,\rm{ fb}^{-1}$ of $e^+e^-$ collisions \cite{Belle-Omega}.

The largest mass gap between the newly observed excited $\Omega_c$ states  and the ground $\Omega_c(2700)$ state is about $400\,\rm{MeV}$, which is large enough
to excite  a light quark-antiquark pair from the QCD vacuum. The observation of the excited $\Omega_c$ states have renewed the interest
in the baryon spectroscopy,  as  the multiquark candidates of those excited states cannot be excluded.
There have been several assignments for those new $\Omega_c$ states, such as the traditional $\Omega_c$ states \cite{S-P-D-Wave},  the compact pentaquark states
\cite{Polyakov-Omega,AnCS-Omega,Pneta-Anisovich},
 molecular pentaquark states \cite{Ping-Omega,Penta-molecule}, dynamically
generated resonances \cite{Penta-Dynamical}.

In Ref.\cite{Polyakov-Omega}, the authors assign the $\Omega_c(3050)$ and $\Omega_c(3119)$  to be the $J^P={\frac{1}{2}}^+$ and ${\frac{3}{2}}^+$ pentaquark states respectively in
the exotic $\overline{15}$ in the chiral quark-soliton model, and take the experimental masses as input parameters to study other pentaquark states.
In Ref.\cite{AnCS-Omega}, An and Chen  study the spectrum of several low-lying $sscq\bar{q}$ pentaquark configurations
using the constituent quark model with  the hyperfine interaction induced by the Goldstone boson exchange.
In Ref.\cite{Pneta-Anisovich}, Anisovich et al study  the mass spectrum of the  $sscq\bar{q}$ pentaquark states with the simple diquark-diquark-antiquark model.

In Ref.\cite{Penta-anti-c}, the authors studied  the scalar-diquark-pseudoscalar-diquark-anti-charmed-quark type pentaquark states $udud\bar{c}$
with the QCD sum rules for the first time by taking into account the vacuum condensates of dimensions $0$, $4$, $6$, $12$ in the operator product expansion,
and obtained a mass around $3.10\,\rm{GeV}$.
In Ref.\cite{Penta-Nielsen},  Albuquerque, Lee and Nielsen  construct both the scalar-diquark-scalar-diquark-antiquark type
and scalar-diquark-pseudoscalar-diquark-antiquark type
currents to study the charmed pentaquark states $udcd\bar{u}$ with $J^P={\frac{1}{2}}^+$ with the QCD sum rules  by
taking into account the vacuum condensates up to dimension $10$
 in the operator product expansion, and obtain the ground state masses $3.21 \pm 0.13\,\rm{GeV}$ and $4.15 \pm 0.11\,\rm{GeV}$, respectively.

In this article, we  will focus on  the  scenario of pentaquark  state interpretation, and study the masses of the $J^P={\frac{3}{2}}^\pm$ charmed pentaquark states
based on the QCD sum rules by taking into account the vacuum condensates up to dimension $13$ in a consistent way,  and revisit the assignments of the new narrow excited $\Omega_c^0$ states.

 The article is arranged as follows:  we derive the QCD sum rules for the masses and pole residues of  the $J^P={\frac{3}{2}}^\pm$ charmed pentaquark
 states in Sect.2;  in Sect.3, we present the numerical results and discussions; and Sect.4 is reserved for our conclusion.

\section{QCD sum rules for  the  ${\frac{3}{2}}^\pm$ charmed pentaquark states}

Firstly, let us write down the two-point correlation functions $\Pi_{\mu\nu}(p)$ in the QCD sum rules,
\begin{eqnarray}
\Pi_{\mu\nu}(p)&=&i\int d^4x e^{ip \cdot x} \langle0|T\left\{J_{\mu}(x)\bar{J}_{\nu}(0)\right\}|0\rangle \, ,
\end{eqnarray}
where  $J_\mu(x)=J_{q^{\prime}q^{\prime\prime}q^{\prime\prime\prime}\bar{q},\mu}(x)$. We choose the axialvector-diquark-scalar-diquark-antiquark type
currents to interpolate the $ J^P={\frac{3}{2}}^{-}$ charmed pentaquark states,
\begin{eqnarray}
J_{q^{\prime}q^{\prime\prime}q^{\prime\prime\prime}\bar{q},\mu}(x)&=&\varepsilon^{ila} \varepsilon^{ijk}\varepsilon^{lmn}
q^{\prime T}_j(x) C\gamma_\mu q^{\prime\prime}_k(x) q^{\prime\prime\prime T}_m(x) C\gamma_5 c_n(x) C\bar{q}^{T}_{a}(x) \, ,
\end{eqnarray}
where  $q^\prime$, $q^{\prime\prime}$, $q^{\prime\prime\prime}=u$, $d$, $s$,  the $i$, $j$, $k$, $l$, $m$, $n$ and $a$ are color indices,
the $C$ is the charge conjugation matrix.

The constituents of the currents $J_{q^{\prime}q^{\prime\prime}q^{\prime\prime\prime}\bar{q},\mu}$ can be divided into two clusters, a diquark
$q^{\prime T}_j C\gamma_\mu q^{\prime\prime}_k$ ($D_i$) plus a triquark $q^{\prime\prime\prime T}_m C\gamma_5 c_n C\bar{q}^{T}_{a}$ ($T_{mna}$).
We take the isospin limit by assuming the $u$ and $d$ quarks have  degenerate masses, and analyze  the isospins of the two clusters,
\begin{eqnarray}
\hat{I}^2 \,\varepsilon_{ijk} q^T_j C\gamma_\mu q^\prime_j &=& 1(1+1)\,\varepsilon_{ijk} q^T_j C\gamma_\mu q^\prime_j \, , \nonumber\\
\hat{I}^2 \,\varepsilon_{ijk} q^T_j C\gamma_\mu s_j &=& \frac{1}{2}\left(\frac{1}{2}+1\right)\,\varepsilon_{ijk} q^T_j C\gamma_\mu s_j \, , \nonumber\\
\hat{I}^2 \,\varepsilon_{ijk} s^T_j C\gamma_\mu s^\prime_j &=& 0(0+1)\,\varepsilon_{ijk} s^T_j C\gamma_\mu s_j \, ,
\end{eqnarray}
\begin{eqnarray}
\hat{I}^2 \, u^T_m C\gamma_5 c_n C \bar{d}^T_a &=& 1\left(1+1\right)\,u^T_m C\gamma_5 c_n C \bar{d}^T_a \, ,  \nonumber\\
\hat{I}^2 \, d^T_m C\gamma_5 c_n C \bar{u}^T_a &=& 1\left(1+1\right)\,d^T_m C\gamma_5 c_n C \bar{u}^T_a \, , \nonumber\\
\hat{I}^2 \,\left[ u^T_m C\gamma_5 c_n C \bar{u}^T_a -d^T_m C\gamma_5 c_n C \bar{d}^T_a\right]&=& 1\left(1+1\right)\,\left[ u^T_m C\gamma_5 c_n C \bar{u}^T_a -d^T_m C\gamma_5 c_n C \bar{d}^T_a\right]\, , \nonumber\\
\hat{I}^2 \,\left[ u^T_m C\gamma_5 c_n C \bar{u}^T_a +d^T_m C\gamma_5 c_n C \bar{d}^T_a\right]&=& 0\left(0+1\right)\,\left[ u^T_m C\gamma_5 c_n C \bar{u}^T_a +d^T_m C\gamma_5 c_n C \bar{d}^T_a\right]\, , \nonumber\\
\hat{I}^2 \, q^T_m C\gamma_5 c_n C \bar{s}^T_a &=& \frac{1}{2}\left(\frac{1}{2}+1\right)\,q^T_m C\gamma_5 c_n C \bar{s}^T_a \, , \nonumber\\
\hat{I}^2 \, s^T_m C\gamma_5 c_n C \bar{q}^T_a &=& \frac{1}{2}\left(\frac{1}{2}+1\right)\,s^T_m C\gamma_5 c_n C \bar{q}^T_a \, , \nonumber\\
\hat{I}^2 \, s^T_m C\gamma_5 c_n C \bar{s}^T_a &=& 0\left(0+1\right)\,s^T_m C\gamma_5 c_n C \bar{s}^T_a \, ,
\end{eqnarray}
where $q$, $q^\prime=u$, $d$, the $\hat{I}^2$ is the isospin operator. The diquark clusters $D_i$ and triquark clusters $T_{mna}$ have the isospins $I=1$, $\frac{1}{2}$ or $0$.
As the color interaction is flavor blinded, we can obtain some mass relations based on the $SU(3)$ breaking effects of the $u$, $d$, $s$ quarks, the mass relation among the diquark clusters $D_i$ is $m_{I=0}-m_{I=\frac{1}{2}}=m_{I=\frac{1}{2}}-m_{I=1}$,
while the mass relation among the triquark clusters $T_{mna}$ is $m_{I=0}-m_{I=\frac{1}{2}}=m_{I=\frac{1}{2}}-m_{I=1}$, if the hidden-flavor (for $u$ and $d$) isospin singlet
$u^T_m C\gamma_5 c_n C \bar{u}^T_a +d^T_m C\gamma_5 c_n C \bar{d}^T_a$ is excluded. On the other hand, the isospin triplet $u^T_m C\gamma_5 c_n C \bar{u}^T_a -d^T_m C\gamma_5 c_n C \bar{d}^T_a$
and isospin singlet $u^T_m C\gamma_5 c_n C \bar{u}^T_a +d^T_m C\gamma_5 c_n C \bar{d}^T_a$ are expected to have degenerate  masses,
which can be inferred from the tiny mass difference between the
vector mesons $\rho^0(770)$ and $\omega(780)$. In fact, if we choose the currents $J_\mu(x)=\bar{u}(x)\gamma_\mu u(x)-\bar{d}(x)\gamma_\mu d(x)$ and
$\bar{u}(x)\gamma_\mu u(x)+\bar{d}(x)\gamma_\mu d(x)$ to interpolate the $\rho^0(770)$ and $\omega(780)$, respectively, we obtain the same  QCD sum rules.

Now we can estimate  the mass differences  among  the pentaquark states $P(D_i,T_{mna})$ according to the $SU(3)$ breaking
effects, or the numbers of the $u$, $d$, $s$ (anti)quarks, i.e. $N_{q}$ and $N_{s}$.
The current $J_{uuu\bar{u},\mu}(x)$ couples potentially to the lowest ground pentaquark state $uuuc\bar{u}$ with $N_{q}=4$
and $N_{s}=0$, while the current $J_{sss\bar{s},\mu}(x)$ couples potentially to the highest ground
pentaquark state $sssc\bar{s}$ with $N_{q}=0$ and $N_{s}=4$. On the other hand,  the currents
$J_{ssu\bar{u},\mu}(x)$, $J_{sus\bar{u},\mu}(x)$, $J_{ssd\bar{d},\mu}(x)$,  $J_{sds\bar{d},\mu}(x)$, $J_{uus\bar{s},\mu}(x)$,  $J_{suu\bar{s},\mu}(x)$,
 $J_{dds\bar{s},\mu}(x)$ and $J_{sdd\bar{s},\mu}(x)$ couple potentially to
the ground state doubly-strange pentaquark states $ssuc\bar{u}$, $susc\bar{u}$, $ssdc\bar{d}$,  $sdsc\bar{d}$ or hidden-strange
pentaquark states  $uusc\bar{s}$,  $suuc\bar{s}$,  $ddsc\bar{s}$,  $sddc\bar{s}$, respectively, which have the quantum numbers $N_{q}=2$ and $N_{s}=2$.
 We expect that the  ground state  pentaquark states with $N_{q}=2$ and $N_{s}=2$ have degenerated masses,
and lie in between the lowest pentaquark state
$uuuc\bar{u}$ and the highest pentaquark state $sssc\bar{s}$.

 In this article, we choose the interpolating currents $J_{uuu\bar{u},\mu}(x)$ and $J_{sss\bar{s},\mu}(x)$ to study the lowest and the
 highest ground states.  Compared to the hidden-charm pentaquark currents and triply-charmed pentaquark currents, it is very difficult to carry out the operator product expansion for the
singly-charmed pentaquark currents.
If the excited $\Omega_c$ states are pentaquark states, they may be couple potentially to the doubly-strange currents $J_{ssu\bar{u},\mu}(x)$, $J_{sus\bar{u},\mu}(x)$,
$J_{ssd\bar{d},\mu}(x)$ or $J_{sds\bar{d},\mu}(x)$. It is even  difficult to carry out the operator product expansion for the currents $J_{ssu\bar{u},\mu}(x)$, $J_{sus\bar{u},\mu}(x)$,
$J_{ssd\bar{d},\mu}(x)$ or $J_{sds\bar{d},\mu}(x)$ due to the $SU(3)$ breaking effects involving the masses and vacuum condensates, such as $m_q$, $m_s$,
$\langle\bar{q}q\rangle$, $\langle\bar{s}s\rangle$, etc.   We can estimate the masses of the doubly-strange pentaquark states  $ssuc\bar{u}$, $susc\bar{u}$, $ssdc\bar{d}$ and $sdsc\bar{d}$
according to the $SU(3)$ breaking effects of the  pentaquark states
$uuuc\bar{u}$ and  $sssc\bar{s}$, and examine the possibility of assigning the new excited $\Omega_c^0$ states as the pentaquark states. If such possibility exists,
we will  study the mass spectrum of the $J^{P}={\frac{1}{2}}^\pm$, ${\frac{3}{2}}^\pm$ and ${\frac{5}{2}}^\pm$ charmed pentaquark states with the QCD sum rules
directly in a systematic way in our next work,
 and examine the assignments of the new excited  $\Omega_c^0$ states exactly.

At the phenomenological  side,  we  insert  a complete set  of intermediate pentaquark  states with the
same quantum numbers as the current operators   $J_{\mu}(x)$ and
$i\gamma_5 J_{\mu}(x)$ into the correlation functions
  $\Pi_{\mu\nu}(p)$ to obtain the hadronic representation
\cite{SVZ79,PRT85}. We isolate the pole terms of the lowest charmed
pentaquark  states with $J^P={\frac{3}{2}}^\pm$,   and obtain the  results:
\begin{eqnarray}
\Pi_{\mu\nu}(p) & = & \left(  \lambda_{-}^2  {\!\not\!{p}+ M_{-} \over M_{-}^{2}-p^{2}  } +\lambda_{+}^2 {\!\not\!{p}- M_{+} \over M_{+}^{2}-p^{2}  } \right)
 \left(- g_{\mu\nu}+\frac{\gamma_\mu\gamma_\nu}{3}+\frac{2p_\mu p_\nu}{3p^2}-\frac{p_\mu\gamma_\nu-p_\nu \gamma_\mu}{3\sqrt{p^2}}
\right) +\cdots  \, ,\nonumber\\
&=&\Pi(p^2)\,\left(- g_{\mu\nu}\right)+\cdots\, .
\end{eqnarray}

The currents    $J_{\mu}(0)$ couple potentially to the spin-parity   ${\frac{3}{2}}^\pm$  charmed pentaquark states \cite{Oka96,WangPc},
\begin{eqnarray}
\langle 0| J_\mu (0)|P_{-}(p)\rangle &=&\lambda_{-} U^{-}_{\mu }(p,s) \, ,\nonumber\\
\langle 0| J_\mu (0)|P_{+}(p)\rangle &=&\lambda_{+} i\gamma_5 U^{+}_{\mu}(p,s) \, ,
\end{eqnarray}
where the $\lambda_{\pm}$  are the pole residues or the current-pentaquark couplings,
the spinors  $U^{\pm}_{\mu}(p,s)$ satisfy the relation,
\begin{eqnarray}
\sum_s U_\mu(p,s) \overline{U}_\nu(p,s)&=&\left(\!\not\!{p}+M_{\pm}\right)
\left(- g_{\mu\nu}+\frac{\gamma_\mu\gamma_\nu}{3}+\frac{2p_\mu p_\nu}{3p^2}-\frac{p_\mu\gamma_\nu-p_\nu \gamma_\mu}{3\sqrt{p^2}}
\right) \,  ,
\end{eqnarray}
and $p^2=M^{2}_{\pm}$ on  mass-shell, the $s$ are the polarizations or spin indexes of the spinors, and should  be distinguished from the $s$ quark or the energy $s$.

We can obtain the hadronic spectral densities at the phenomenological  side through dispersion relation,
\begin{eqnarray}
\frac{{\rm Im}\Pi(s)}{\pi}&=&\!\not\!{p} \left[\lambda_{-}^2 \delta\left(s-M_{-}^2\right)+\lambda_{+}^2 \delta\left(s-M_{+}^2\right)\right] +\left[M_{-}\lambda_{-}^2 \delta\left(s-M_{-}^2\right)-M_{+}\lambda_{+}^2 \delta\left(s-M_{+}^2\right)\right]  \nonumber\\
&=&\!\not\!{p}\, \rho^1_{H}(s)+\rho^0_{H}(s) \, ,
\end{eqnarray}
where the subscript $H$ denotes  the hadron (or phenomenological) side,
then we introduce the weight function $\exp\left(-\frac{s}{T^2}\right)$  to obtain the QCD sum rules at the hadron side,
\begin{eqnarray}
\int_{m_c^2}^{s_0}ds \left[\sqrt{s}\rho^1_{H}(s)+\rho^0_{H}(s)\right]\exp\left( -\frac{s}{T^2}\right)
&=&2M_{-}\lambda_{-}^2\exp\left( -\frac{M_{-}^2}{T^2}\right) \, ,  \\
\int_{m_c^2}^{s_0}ds \left[\sqrt{s}\rho^1_{H}(s)-\rho^0_{H}(s)\right]\exp\left( -\frac{s}{T^2}\right)
&=&2M_{+}\lambda_{+}^2\exp\left( -\frac{M_{+}^2}{T^2}\right) \, ,
\end{eqnarray}
where the $s_0$ are the continuum threshold parameters and the $T^2$ are the Borel parameters \cite{WangPc}. We separate
the contributions of the negative parity and positive   parity charmed pentaquark states    unambiguously according to Eqs.(9-11).

Now we briefly outline  the operator product expansion for the correlation function $ \Pi_{sss\bar{s},\mu\nu}(p)$, the correlation function $ \Pi_{uuu\bar{u},\mu\nu}(p)$ can be
obtained with a simple replacement. Firstly,  we contract the   $s$ and $c$ quark fields in the correlation function $\Pi_{sss\bar{s},\mu\nu}(p)$    with Wick theorem,
\begin{eqnarray}
\Pi_{sss\bar{s},\mu\nu}(p)&=&i\,\varepsilon^{ila}\varepsilon^{ijk}\varepsilon^{lmn}\varepsilon^{i^{\prime}l^{\prime}a^{\prime}}\varepsilon^{i^{\prime}j^{\prime}k^{\prime}}
\varepsilon^{l^{\prime}m^{\prime}n^{\prime}}\int d^4x e^{ip\cdot x}C S_{a^{\prime}a}^T(-x)C \nonumber\\
&&\Big\{ 2  Tr\left[\gamma_\mu S_{kk^\prime}(x) \gamma_\nu C S^{T}_{jj^\prime}(x)C\right] \,Tr\left[\gamma_5 C_{nn^\prime}(x) \gamma_5 C S^{T}_{mm^\prime}(x)C\right]   \nonumber\\
&&-4  Tr \left[\gamma_\mu S_{kk^\prime}(x) \gamma_\nu C S^{T}_{mj^\prime}(x)C \gamma_5C_{nn^\prime}(x) \gamma_5 C S^{T}_{jm^\prime}(x)C\right]  \Big\} \, ,
\end{eqnarray}
where the  $S_{ij}(x)$ and $C_{ij}(x)$ are the full  $s$ and $c$ quark propagators, respectively \cite{PRT85,Pascual-1984},
 \begin{eqnarray}
S_{ij}(x)&=& \frac{i\delta_{ij}\!\not\!{x}}{ 2\pi^2x^4}-\frac{\delta_{ij}m_s}{4\pi^2x^2}-\frac{\delta_{ij}\langle
\bar{s}s\rangle}{12} +\frac{i\delta_{ij}\!\not\!{x}m_s\langle\bar{s}s\rangle}{48}-\frac{\delta_{ij}x^2\langle \bar{s}g_s\sigma Gs\rangle}{192}\nonumber\\
&&+\frac{i\delta_{ij}x^2\!\not\!{x} m_s\langle \bar{s}g_s\sigma Gs\rangle }{1152} -\frac{ig_s G^{a}_{\alpha\beta}t^a_{ij}(\!\not\!{x}
\sigma^{\alpha\beta}+\sigma^{\alpha\beta} \!\not\!{x})}{32\pi^2x^2} -\frac{1}{8}\langle\bar{s}_j\sigma^{\mu\nu}s_i \rangle \sigma_{\mu\nu} +\cdots \, , \nonumber \\
\end{eqnarray}
\begin{eqnarray}
C_{ij}(x)&=&\frac{i}{(2\pi)^4}\int d^4k e^{-ik \cdot x} \left\{
\frac{\delta_{ij}}{\!\not\!{k}-m_c}
-\frac{g_sG^n_{\alpha\beta}t^n_{ij}}{4}\frac{\sigma^{\alpha\beta}(\!\not\!{k}+m_c)+(\!\not\!{k}+m_c)
\sigma^{\alpha\beta}}{(k^2-m_c^2)^2}+\cdots\right\} \, , \nonumber\\
\end{eqnarray}
 $t^n=\frac{\lambda^n}{2}$, the $\lambda^n$ is the Gell-Mann matrix. In Eq.(13), we add the term
  $\langle\bar{s}_j\sigma_{\mu\nu}s_i \rangle$  originates  from the Fierz re-ordering   of the
   $\langle s_i \bar{s}_j\rangle$ to  absorb the gluons  emitted from other quark lines to  extract the mixed condensate     $\langle\bar{s}g_s\sigma G s\rangle$.
    The term $-\frac{1}{8}\langle\bar{s}_j\sigma^{\mu\nu}s_i \rangle \sigma_{\mu\nu}$ was introduced in Ref.\cite{WangTetraquark}.
We compute  the integrals both in the coordinate space and momentum space  to obtain the correlation function $\Pi(p^2)$ at the quark level, then obtain  the QCD spectral densities
 through  dispersion relation,
\begin{eqnarray}
\frac{{\rm Im}\Pi(s)}{\pi}&=&\!\not\!{p}\, \rho^1_{QCD}(s)+\rho^0_{QCD}(s) \, ,
\end{eqnarray}
the $\rho^1_{QCD}(s)$ and  $\rho^0_{QCD}(s)$ are the QCD spectral densities.

We  take the quark-hadron duality,  introduce the continuum threshold parameters  $s_0$ and the weight function $\exp\left(-\frac{s}{T^2}\right)$  to obtain  the  QCD sum rules:
\begin{eqnarray}
\int_{m_c^2}^{s_0}ds \left[\sqrt{s}\rho^1_{H}(s)+\rho^0_{H}(s)\right]\exp\left( -\frac{s}{T^2}\right)
&=& \int_{m_c^2}^{s_0}ds \left[\sqrt{s}\rho^1_{QCD}(s)+\rho^0_{QCD}(s)\right]\exp\left( -\frac{s}{T^2}\right)\, ,  \nonumber\\
\int_{m_c^2}^{s_0}ds \left[\sqrt{s}\rho^1_{H}(s)-\rho^0_{H}(s)\right]\exp\left( -\frac{s}{T^2}\right)
&=& \int_{m_c^2}^{s_0}ds \left[\sqrt{s}\rho^1_{QCD}(s)-\rho^0_{QCD}(s)\right]\exp\left( -\frac{s}{T^2}\right)\,  ,\nonumber\\
\end{eqnarray}
 where $\rho^1_{QCD}(s)=\rho^1(s)+\rho^1_F(s)$ and $\rho^0_{QCD}(s)=\rho^0(s)+\rho^0_F(s)$,
 \begin{eqnarray}
\rho^{1}(s)&=&\frac{1}{7372800\pi^8}\int_{x_i}^1dx \,x\bar{x}^5(x+5)(s-\widetilde{m}_c^2)^5  +\frac{m_s m_c}{1474560\pi^8}\int_{x_i}^1dx \,\bar{x}^5(x+5)(s-\widetilde{m}_c^2)^4 \nonumber\\
&&-\frac{m_c\langle \bar{s}s\rangle}{46080\pi^6}\int_{x_i}^1dx \,\bar{x}^4(x+4)(s-\widetilde{m}_c^2)^3 -\frac{m_s\langle \bar{s}s\rangle}{4608\pi^6}\int_{x_i}^1dx \,x\bar{x}^4(s-\widetilde{m}_c^2)^3 \nonumber\\
&&+\frac{m_c\langle \bar{s}g_s\sigma Gs\rangle}{12288\pi^6}\int_{x_i}^1dx \,\bar{x}^3(x+3)(s-\widetilde{m}_c^2)^2 +\frac{m_s\langle \bar{s}g_s\sigma Gs\rangle}{4608\pi^6}\int_{x_i}^1dx \,x\bar{x}^2(5-2x)(s-\widetilde{m}_c^2)^2 \nonumber\\
&&+\frac{\langle \bar{s} s\rangle^2}{192\pi^4}\int_{x_i}^1dx \,x\bar{x}^2(s-\widetilde{m}_c^2)^2 +\frac{m_s m_c\langle \bar{s} s\rangle^2}{192\pi^4}\int_{x_i}^1dx \,\bar{x}^2(4-x)(s-\widetilde{m}_c^2) \nonumber\\
&&-\frac{\langle \bar{s} s\rangle \langle\bar{s}g_s\sigma Gs\rangle }{96\pi^4}\int_{x_i}^1dx \,x\bar{x} (s-\widetilde{m}_c^2)  +\frac{m_s m_c\langle \bar{s} s\rangle \langle\bar{s}g_s\sigma Gs\rangle }{768\pi^4}\int_{x_i}^1dx \,\bar{x}(5x-19)   \nonumber\\
&&-\frac{m_c \langle \bar{s} s\rangle^3 }{36\pi^2}\int_{x_i}^1dx \,\bar{x}    +\frac{m_s \langle \bar{s} s\rangle^3 }{36\pi^2}\int_{x_i}^1dx \,x
 +\frac{  \langle\bar{s}g_s\sigma Gs\rangle^2 }{768\pi^4}\int_{x_i}^1dx \, x     \nonumber\\
&&+\frac{ m_s m_c \langle\bar{s}g_s\sigma Gs\rangle^2 }{768\pi^4}\int_{x_i}^1dx \, (3-x)\delta (s-\widetilde{m}_c^2)     +\frac{m_c \langle \bar{s} s\rangle^2 \langle\bar{s}g_s\sigma Gs\rangle }{48\pi^2}\int_{x_i}^1dx \,\delta (s-\widetilde{m}_c^2)  \nonumber\\
&&-\frac{m_s \langle \bar{s} s\rangle^2 \langle\bar{s}g_s\sigma Gs\rangle }{54\pi^2} \,\delta (s-m_c^2)  -\frac{m_s m_c \langle \bar{s} s\rangle^4  }{54T^2}  \,\delta (s-m_c^2)  \nonumber\\
&&-\frac{m_c \langle \bar{s} s\rangle  \langle\bar{s}g_s\sigma Gs\rangle^2 }{192\pi^2 T^2}  \,\delta (s-m_c^2)
+\frac{7m_s \langle \bar{s} s\rangle  \langle\bar{s}g_s\sigma Gs\rangle^2 }{1728\pi^2 T^2}\left(1+\frac{s}{T^2} \right)  \,\delta (s-m_c^2)  \, ,
\end{eqnarray}

\begin{eqnarray}
\rho^{0}(s)&=&\frac{m_s}{2457600\pi^8}\int_{x_i}^1dx \,x\bar{x}^5(x+3)(s-\widetilde{m}_c^2)^5 +\frac{m_s}{1474560\pi^8}\int_{x_i}^1dx \,x\bar{x}^5(x+5)s(s-\widetilde{m}_c^2)^4 \nonumber\\
&&-\frac{\langle \bar{s}s\rangle}{184320\pi^6}\int_{x_i}^1dx \,x\bar{x}^4(3x+7)(s-\widetilde{m}_c^2)^4 -\frac{\langle \bar{s}s\rangle}{46080\pi^6}\int_{x_i}^1dx \,x\bar{x}^4(x+4)s(s-\widetilde{m}_c^2)^3 \nonumber\\
&&-\frac{m_s m_c\langle \bar{s}s\rangle}{4608\pi^6}\int_{x_i}^1dx \,\bar{x}^3(x+3)(s-\widetilde{m}_c^2)^3
+\frac{\langle \bar{s}g_s\sigma Gs\rangle}{36864\pi^6}\int_{x_i}^1dx \,x\bar{x}^3(3x+5)(s-\widetilde{m}_c^2)^3 \nonumber\\
&&+\frac{\langle \bar{s}g_s\sigma Gs\rangle}{12288\pi^6}\int_{x_i}^1dx \,x\bar{x}^3(x+3)s(s-\widetilde{m}_c^2)^2  +\frac{m_s m_c \langle \bar{s}g_s\sigma Gs\rangle}{1536\pi^6}\int_{x_i}^1dx \,\bar{x}^2(x+2) (s-\widetilde{m}_c^2)^2 \nonumber\\
&&+\frac{m_c \langle \bar{s} s\rangle^2}{576\pi^4}\int_{x_i}^1dx \,\bar{x}^2(x+2)(s-\widetilde{m}_c^2)^2
+\frac{m_s \langle \bar{s} s\rangle^2}{128\pi^4}\int_{x_i}^1dx \,x\bar{x}^2(3-x)(s-\widetilde{m}_c^2)^2 \nonumber\\
&&+\frac{m_s \langle \bar{s} s\rangle^2}{192\pi^4}\int_{x_i}^1dx \,x\bar{x}^2(4-x)s(s-\widetilde{m}_c^2)  -\frac{m_c \langle \bar{s} s\rangle \langle\bar{s}g_s\sigma Gs\rangle }{192\pi^4}\int_{x_i}^1dx \,\bar{x}(x+1) (s-\widetilde{m}_c^2)  \nonumber\\
&&+\frac{m_s \langle \bar{s} s\rangle \langle\bar{s}g_s\sigma Gs\rangle }{768\pi^4}\int_{x_i}^1dx \,x\bar{x}(15x-43) (s-\widetilde{m}_c^2)  +\frac{m_s \langle \bar{s} s\rangle \langle\bar{s}g_s\sigma Gs\rangle }{768\pi^4}\int_{x_i}^1dx \,x\bar{x}(5x-19) s  \nonumber\\
&&-\frac{  \langle \bar{s} s\rangle^3 }{36\pi^2}\int_{x_i}^1dx \,x\bar{x}(3s-2\widetilde{m}_c^2)
+\frac{ m_s m_c \langle \bar{s} s\rangle^3 }{36\pi^2}\int_{x_i}^1dx \,(x-4)   +\frac{ m_c \langle\bar{s}g_s\sigma Gs\rangle^2 }{768\pi^4}\int_{x_i}^1dx \, x     \nonumber\\
&&+\frac{ m_s \langle\bar{s}g_s\sigma Gs\rangle^2 }{768\pi^4}\int_{x_i}^1dx \,x (7-3x)     +\frac{ m_s \langle\bar{s}g_s\sigma Gs\rangle^2 }{768\pi^4}\int_{x_i}^1dx \,x (3-x) s  \delta(s-\widetilde{m}_c^2)   \nonumber\\
&&+\frac{ \langle \bar{s} s\rangle^2 \langle\bar{s}g_s\sigma Gs\rangle }{24\pi^2}\int_{x_i}^1dx \,x\left[1+\frac{s}{2} \delta (s-\widetilde{m}_c^2) \right]
+\frac{m_s m_c \langle \bar{s} s\rangle^2 \langle\bar{s}g_s\sigma Gs\rangle }{54\pi^2}\int_{x_i}^1dx \,  \delta (s-\widetilde{m}_c^2)   \nonumber\\
&&+\frac{7m_s m_c \langle \bar{s} s\rangle^2 \langle\bar{s}g_s\sigma Gs\rangle }{108\pi^2} \,  \delta (s-m_c^2)   +\frac{ m_c \langle \bar{s} s\rangle^4  }{27}  \,\delta (s-m_c^2) -\frac{ m_s \langle \bar{s} s\rangle^4  }{54} \left(1+\frac{s}{T^2} \right) \,\delta (s-m_c^2)  \nonumber\\
&&-\frac{ \langle \bar{s} s\rangle  \langle\bar{s}g_s\sigma Gs\rangle^2 }{192\pi^2 }\left( 1+\frac{s}{T^2}\right)  \,\delta (s-m_c^2)  -\frac{ 29m_s m_c^3\langle \bar{s} s\rangle  \langle\bar{s}g_s\sigma Gs\rangle^2 }{1728\pi^2T^4 }   \,\delta (s-m_c^2)  \nonumber\\
&&-\frac{ 7m_s m_c\langle \bar{s} s\rangle  \langle\bar{s}g_s\sigma Gs\rangle^2 }{1728\pi^2T^2 }   \,\delta (s-m_c^2)  \, ,
\end{eqnarray}

\begin{eqnarray}
\rho^{1}_{F}(s)&=&-\frac{m_c\langle \bar{s}g_s\sigma Gs\rangle}{61440\pi^6}\int_{x_i}^1dx \frac{\bar{x}^4(x+4)}{x}(s-\widetilde{m}_c^2)^2 +\frac{m_c\langle \bar{s}g_s\sigma Gs\rangle}{491520\pi^6}\int_{x_i}^1dx \frac{\bar{x}^4(2x+3)}{x}(s-\widetilde{m}_c^2)^2 \nonumber\\
&&+\frac{m_s\langle \bar{s}g_s\sigma Gs\rangle}{49152\pi^6}\int_{x_i}^1dx\, \bar{x}^3(x+1)\,(s-\widetilde{m}_c^2)^2 +\frac{m_c\langle \bar{s}g_s\sigma Gs\rangle}{32768\pi^6}\int_{x_i}^1dx\, \bar{x}^3(x+3)\,(s-\widetilde{m}_c^2)^2 \nonumber\\
&&+\frac{3m_s\langle \bar{s}g_s\sigma Gs\rangle}{8192\pi^6}\int_{x_i}^1dx\, x\bar{x}^2\,(s-\widetilde{m}_c^2)^2 +\frac{m_s m_c \langle \bar{s} s\rangle \langle \bar{s}g_s\sigma Gs\rangle}{768\pi^4}\int_{x_i}^1dx \frac{\bar{x}^2(2-x)}{x}  \nonumber\\
&&-\frac{  \langle \bar{s} s\rangle \langle \bar{s}g_s\sigma Gs\rangle}{9216\pi^4}\int_{x_i}^1dx \,\bar{x}^2(2x+1)\, (s-\widetilde{m}_c^2) +\frac{ m_s m_c \langle \bar{s} s\rangle \langle \bar{s}g_s\sigma Gs\rangle}{6144\pi^4}\int_{x_i}^1dx \,\frac{\bar{x}^2(2x+1)}{x} \nonumber\\
&&-\frac{  \langle \bar{s} s\rangle \langle \bar{s}g_s\sigma Gs\rangle}{512\pi^4}\int_{x_i}^1dx \,x\bar{x} \, (s-\widetilde{m}_c^2) +\frac{ m_s m_c \langle \bar{s} s\rangle \langle \bar{s}g_s\sigma Gs\rangle}{1024\pi^4}\int_{x_i}^1dx \,\bar{x}(x+1)\nonumber\\
&&+\frac{m_s m_c  \langle \bar{s}g_s\sigma Gs\rangle^2}{1536\pi^4}\int_{x_i}^1dx \frac{\bar{x}(x-3)}{x} \delta(s-\widetilde{m}_c^2) +\frac{   \langle \bar{s}g_s\sigma Gs\rangle^2}{27648\pi^4}\int_{x_i}^1dx \,x\bar{x}   \nonumber\\
&&+\frac{   \langle \bar{s}g_s\sigma Gs\rangle^2}{6144\pi^4}\int_{x_i}^1dx \,x\bar{x}   -\frac{ m_s m_c  \langle \bar{s}g_s\sigma Gs\rangle^2}{6144\pi^4}\int_{x_i}^1dx \,\bar{x} \, \delta(s-\widetilde{m}_c^2) \nonumber\\
&&+\frac{   \langle \bar{s}g_s\sigma Gs\rangle^2}{2048\pi^4}\int_{x_i}^1dx \,x -\frac{ m_s m_c  \langle \bar{s}g_s\sigma Gs\rangle^2}{3072\pi^4}\int_{x_i}^1dx \,x \, \delta(s-\widetilde{m}_c^2) \nonumber\\
&&-\frac{ m_c  \langle \bar{s} s\rangle^2 \langle \bar{s}g_s\sigma Gs\rangle }{144\pi^2}\int_{x_i}^1dx \frac{\bar{x} }{x} \delta(s-\widetilde{m}_c^2) +\frac{ m_s  \langle \bar{s} s\rangle^2 \langle \bar{s}g_s\sigma Gs\rangle }{1152\pi^2}\int_{x_i}^1dx \,(1-2x)\, \delta(s-\widetilde{m}_c^2) \nonumber\\
&&-\frac{ m_s  \langle \bar{s} s\rangle^2 \langle \bar{s}g_s\sigma Gs\rangle }{768\pi^2}\, \delta(s-m_c^2) +\frac{ m_c  \langle \bar{s} s\rangle  \langle \bar{s}g_s\sigma Gs\rangle^2 }{288\pi^2 T^2}\int_{x_i}^1dx \frac{1 }{x} \delta(s-\widetilde{m}_c^2) \nonumber\\
&&-\frac{ m_s  \langle \bar{s} s\rangle  \langle \bar{s}g_s\sigma Gs\rangle^2 }{20736\pi^2 T^2}\int_{x_i}^1dx  \, \delta(s-\widetilde{m}_c^2) +\frac{ m_s  \langle \bar{s} s\rangle  \langle \bar{s}g_s\sigma Gs\rangle^2 }{41472\pi^2 T^2}   \, \delta(s-m_c^2) \nonumber\\
&&+\frac{5 m_s  \langle \bar{s} s\rangle  \langle \bar{s}g_s\sigma Gs\rangle^2 }{13824\pi^2 T^2}    \, \delta(s-m_c^2) -\frac{5 m_s  \langle \bar{s} s\rangle  \langle \bar{s}g_s\sigma Gs\rangle^2 }{6912\pi^2 T^2}  \int_{x_i}^1dx   \, \delta(s-\widetilde{m}_c^2) \nonumber\\
&&+\frac{5 m_s  \langle \bar{s} s\rangle  \langle \bar{s}g_s\sigma Gs\rangle^2 }{9216\pi^2 T^2}     \,\left(1+\frac{s}{T^2} \right) \delta(s-m_c^2) \, ,
\end{eqnarray}

\begin{eqnarray}
\rho^{0}_{F}(s)&=&-\frac{m_s m_c\langle \bar{s}g_s\sigma Gs\rangle}{12288\pi^6}\int_{x_i}^1dx \frac{\bar{x}^3(x+3)}{x}(s-\widetilde{m}_c^2)^2  +\frac{m_s m_c\langle \bar{s}g_s\sigma Gs\rangle}{49152\pi^6}\int_{x_i}^1dx \frac{\bar{x}^3(x+1)}{x}(s-\widetilde{m}_c^2)^2 \nonumber\\
&&+\frac{m_s m_c\langle \bar{s}g_s\sigma Gs\rangle}{8192\pi^6}\int_{x_i}^1dx \, \bar{x}^2(x+2)\,(s-\widetilde{m}_c^2)^2 +\frac{ m_c\langle \bar{s} s\rangle \langle \bar{s}g_s\sigma Gs\rangle}{1152\pi^4}\int_{x_i}^1dx \frac{\bar{x}^2(x+2)}{x}(s-\widetilde{m}_c^2) \nonumber\\
&&-\frac{ m_c\langle \bar{s} s\rangle \langle \bar{s}g_s\sigma Gs\rangle}{9216\pi^4}\int_{x_i}^1dx \frac{\bar{x}^2(2x+1)}{x}(s-\widetilde{m}_c^2) -\frac{ m_s\langle \bar{s} s\rangle \langle \bar{s}g_s\sigma Gs\rangle}{1536\pi^4}\int_{x_i}^1dx \,\bar{x}^2\,s \nonumber\\
&&-\frac{  m_c\langle \bar{s} s\rangle\langle \bar{s}g_s\sigma Gs\rangle}{1024\pi^6}\int_{x_i}^1dx \,\bar{x} (x+1)\,(s-\widetilde{m}_c^2)  -\frac{  m_s\langle \bar{s} s\rangle\langle \bar{s}g_s\sigma Gs\rangle}{256\pi^6}\int_{x_i}^1dx \,x\bar{x}  \,(3s-2\widetilde{m}_c^2)  \nonumber\\
&&-\frac{ m_c \langle \bar{s}g_s\sigma Gs\rangle^2}{1536\pi^4}\int_{x_i}^1dx \frac{\bar{x}(x+1)}{x}  +\frac{ m_s \langle \bar{s}g_s\sigma Gs\rangle^2}{27648\pi^4}\int_{x_i}^1dx \,\bar{x}\,s\, \delta(s-\widetilde{m}_c^2) \nonumber\\
&&+\frac{ m_c \langle \bar{s}g_s\sigma Gs\rangle^2}{6144\pi^4}\int_{x_i}^1dx\, \bar{x}  +\frac{ m_s \langle \bar{s}g_s\sigma Gs\rangle^2}{3072\pi^4}\int_{x_i}^1dx\, \bar{x} \,s\, \delta(s-\widetilde{m}_c^2) \nonumber\\
&&+\frac{ m_c \langle \bar{s}g_s\sigma Gs\rangle^2}{2048\pi^4}\int_{x_i}^1dx\, x +\frac{ m_s \langle \bar{s}g_s\sigma Gs\rangle^2}{512\pi^4}\int_{x_i}^1dx\, x \,\left[1+\frac{s}{2}\, \delta(s-\widetilde{m}_c^2) \right]\nonumber\\
&&+\frac{ m_s m_c \langle \bar{s} s\rangle^2\langle \bar{s}g_s\sigma Gs\rangle}{144\pi^2}\int_{x_i}^1dx \frac{x-3}{x} \delta(s-\widetilde{m}_c^2) +\frac{   \langle \bar{s} s\rangle^2\langle \bar{s}g_s\sigma Gs\rangle}{1152\pi^2}\int_{x_i}^1dx \,\bar{x}\,s\, \delta(s-\widetilde{m}_c^2) \nonumber\\
&&+\frac{ m_s m_c \langle \bar{s} s\rangle^2\langle \bar{s}g_s\sigma Gs\rangle}{1152\pi^2}\int_{x_i}^1dx \frac{1-2x}{x} \delta(s-\widetilde{m}_c^2) +\frac{  \langle \bar{s} s\rangle^2\langle \bar{s}g_s\sigma Gs\rangle}{192\pi^2}\int_{x_i}^1dx\,x\, \left[1+\frac{s}{2} \delta(s-\widetilde{m}_c^2) \right]\nonumber\\
&&-\frac{ m_s m_c \langle \bar{s} s\rangle^2\langle \bar{s}g_s\sigma Gs\rangle}{768\pi^2}  \delta(s-m_c^2) +\frac{ m_s m_c \langle \bar{s} s\rangle^2\langle \bar{s}g_s\sigma Gs\rangle}{768\pi^2} \int_{x_i}^1dx\, \delta(s-\widetilde{m}_c^2) \nonumber\\
&&+\frac{5 m_s m_c \langle \bar{s} s\rangle\langle \bar{s}g_s\sigma Gs\rangle^2}{1728\pi^2T^2}\int_{x_i}^1dx \frac{1}{x} \delta(s-\widetilde{m}_c^2)
+\frac{13 m_s m_c \langle \bar{s} s\rangle\langle \bar{s}g_s\sigma Gs\rangle^2}{1728\pi^2T^2}  \delta(s-m_c^2) \nonumber\\
&&+\frac{ \langle \bar{s} s\rangle\langle \bar{s}g_s\sigma Gs\rangle^2}{20736\pi^2 }\int_{x_i}^1dx \,\left(1-\frac{s}{T^2} \right) \delta(s-\widetilde{m}_c^2) +\frac{ \langle \bar{s} s\rangle\langle \bar{s}g_s\sigma Gs\rangle^2}{2304\pi^2 }\int_{x_i}^1dx \,\left(1-\frac{s}{T^2} \right) \delta(s-\widetilde{m}_c^2) \nonumber\\
&&+\frac{ 5m_s m_c\langle \bar{s} s\rangle\langle \bar{s}g_s\sigma Gs\rangle^2}{13824\pi^2 T^2} \delta(s-m_c^2) -\frac{ 5m_s m_c\langle \bar{s} s\rangle\langle \bar{s}g_s\sigma Gs\rangle^2}{6912\pi^2 T^2} \int_{x_i}^1dx \, \frac{1}{x}\delta(s-\widetilde{m}_c^2) \nonumber\\
&&-\frac{ \langle \bar{s} s\rangle\langle \bar{s}g_s\sigma Gs\rangle^2}{768\pi^2 }\int_{x_i}^1dx \,\left(1+\frac{s}{T^2} \right) \delta(s-\widetilde{m}_c^2) +\frac{5m_s m_c^3 \langle \bar{s} s\rangle\langle \bar{s}g_s\sigma Gs\rangle^2}{9216\pi^2T^4 }\, \delta(s-m_c^2) \nonumber\\
&&-\frac{5m_s m_c \langle \bar{s} s\rangle\langle \bar{s}g_s\sigma Gs\rangle^2}{9216\pi^2T^2 }\, \delta(s-m_c^2) \, ,
\end{eqnarray}
$\widetilde{m}_c^2=\frac{m_c^2}{x}$, $x_i=\frac{m_c^2}{s}$, $\bar{x}=1-x$. With a simple replacement $m_s\to 0$, $\langle\bar{s}s\rangle \to \langle\bar{q}q\rangle$,
$\langle\bar{s}g_s\sigma Gs\rangle \to \langle\bar{q}g_s\sigma Gq\rangle$, we can obtain the QCD spectral densities of the $uuu\bar{u}c$ pentaquark state.
The components $\rho^1_F(s)$ and $\rho^0_F(s)$ denote the contributions  involving the term,
\begin{eqnarray}
-\frac{1}{8}\langle\bar{s}_j\sigma^{\mu\nu}s_i \rangle \sigma_{\mu\nu}  \, ,
\end{eqnarray}
 in the full
$s$-quark propagator. From the spectral densities $\rho^1_F(s)$ and $\rho^0_F(s)$, we can see that there are contributions of the vacuum condensates of dimensions $5$, $8$,
$10$, $11$ and $13$, which play an important role in determining  the Borel windows therefore the predicted masses.

In this article, we carry out the operator product expansion to the vacuum condensates  up to dimension-$13$ and assume vacuum saturation for the
 higher dimension vacuum condensates. We take the truncations $n\leq 13$ and $k\leq 1$ in a consistent way,
the operators of the orders $\mathcal{O}( \alpha_s^{k})$ with $k> 1$ are  discarded. For example,
the vacuum condensates $\langle g_s^3 GGG\rangle$, $\langle \frac{\alpha_s GG}{\pi}\rangle^2$,
 $\langle \frac{\alpha_s GG}{\pi}\rangle\langle \bar{s} g_s \sigma Gs\rangle$, $\langle \frac{\alpha_s GG}{\pi}\rangle^2 \langle\bar{s}s\rangle$ have the dimensions
 $6$, $8$, $9$, $11$, respectively,  but they are   the vacuum expectations
of the operators of the order    $\mathcal{O}( \alpha_s^{3/2})$, $\mathcal{O}(\alpha_s^2)$, $\mathcal{O}( \alpha_s^{3/2})$, $\mathcal{O}(\alpha_s^2)$, respectively, and
are discarded.
The vacuum  condensates $\langle \frac{\alpha_s}{\pi}GG\rangle$, $\langle \bar{s}s\rangle\langle \frac{\alpha_s}{\pi}GG\rangle$,
$\langle \bar{s}s\rangle^2\langle \frac{\alpha_s}{\pi}GG\rangle$ and $\langle \bar{s}s\rangle^3\langle \frac{\alpha_s}{\pi}GG\rangle$ are the vacuum expectations
of the operators of the order
$\mathcal{O}(\alpha_s)$, and they are neglected due to the small contributions of the  gluon condensates for the hidden-charm (hidden-bottom) tetraquark states and hidden-charm
pentaquark states \cite{WangPc,WangTetraquark}.

The QCD sum rules can be written more explicitly,
\begin{eqnarray}
2M_{-}\lambda_{-}^2\exp\left( -\frac{M_{-}^2}{T^2}\right)
&=& \int_{m_c^2}^{s_0}ds \left[\sqrt{s}\rho^1_{QCD}(s)+\rho^0_{QCD}(s)\right]\exp\left( -\frac{s}{T^2}\right)\, ,  \\
2M_{+}\lambda_{+}^2\exp\left( -\frac{M_{+}^2}{T^2}\right)
&=& \int_{m_c^2}^{s_0}ds \left[\sqrt{s}\rho^1_{QCD}(s)-\rho^0_{QCD}(s)\right] \exp\left( -\frac{s}{T^2}\right)\, .
\end{eqnarray}
The contributions of the negative  parity and  positive parity charmed pentaquark  states are separated explicitly.

We derive    Eqs.(22-23) with respect to  $\tau=\frac{1}{T^2}$, then eliminate the
 pole residues $\lambda_{\pm}$ and obtain the QCD sum rules for
 the masses of the charmed pentaquark states,
 \begin{eqnarray}
 M^2_{-} &=& \frac{-\frac{d}{d\tau}\int_{m_c^2}^{s_0}ds \left[\sqrt{s}\rho_{QCD}^1(s)+\rho_{QCD}^0(s)\right]\exp\left( -s\tau\right)}
 {\int_{m_c^2}^{s_0}ds \left[\sqrt{s}\rho_{QCD}^1(s)+\rho_{QCD}^0(s)\right]\exp\left( -s\tau\right)}\, ,\\
 M^2_{+} &=& \frac{-\frac{d}{d\tau}\int_{m_c^2}^{s_0}ds \left[\sqrt{s}\rho_{QCD}^1(s)-\rho_{QCD}^0(s)\right]\exp\left( -s\tau\right)}
 {\int_{m_c^2}^{s_0}ds \left[\sqrt{s}\rho_{QCD}^1(s)-\rho_{QCD}^0(s)\right]\exp\left( -s\tau\right)}\, .
\end{eqnarray}

\section{Numerical results and discussions}
We take  the standard values of the vacuum condensates $\langle
\bar{q}q \rangle=-(0.24\pm 0.01\, \rm{GeV})^3$,   $\langle
\bar{q}g_s\sigma G q \rangle=m_0^2\langle \bar{q}q \rangle$, $\langle\bar{s}g_s\sigma G s \rangle=m_0^2\langle \bar{s}s \rangle$,
$m_0^2=(0.8 \pm 0.1)\,\rm{GeV}^2$, $\langle\bar{s}s \rangle=(0.8\pm0.1)\langle\bar{q}q \rangle$,   $\langle \frac{\alpha_s
GG}{\pi}\rangle=(0.33\,\rm{GeV})^4 $    at the energy scale  $\mu=1\, \rm{GeV}$
\cite{SVZ79,PRT85,ColangeloReview}, and choose the $\overline{MS}$ masses  $m_{c}(m_c)=(1.28\pm0.03)\,\rm{GeV}$ and $m_s(\mu=2\,\rm{GeV})=0.096^{+0.008}_{-0.004}\,\rm{GeV}$
 from the Particle Data Group \cite{PDG}.
Furthermore, we take into account the energy-scale dependence of  the input parameters,
\begin{eqnarray}
\langle\bar{q}q \rangle(\mu)&=&\langle\bar{q}q \rangle(Q)\left[\frac{\alpha_{s}(Q)}{\alpha_{s}(\mu)}\right]^{\frac{12}{25}}\, , \nonumber\\
 \langle\bar{s}s \rangle(\mu)&=&\langle\bar{s}s \rangle(Q)\left[\frac{\alpha_{s}(Q)}{\alpha_{s}(\mu)}\right]^{\frac{12}{25}}\, , \nonumber\\
 \langle\bar{q}g_s \sigma Gq \rangle(\mu)&=&\langle\bar{q}g_s \sigma Gq \rangle(Q)\left[\frac{\alpha_{s}(Q)}{\alpha_{s}(\mu)}\right]^{\frac{2}{25}}\, , \nonumber\\
 \langle\bar{s}g_s \sigma Gs \rangle(\mu)&=&\langle\bar{s}g_s \sigma Gs \rangle(Q)\left[\frac{\alpha_{s}(Q)}{\alpha_{s}(\mu)}\right]^{\frac{2}{25}}\, , \nonumber\\
m_c(\mu)&=&m_c(m_c)\left[\frac{\alpha_{s}(\mu)}{\alpha_{s}(m_c)}\right]^{\frac{12}{25}} \, ,\nonumber\\
m_s(\mu)&=&m_s({\rm 2GeV} )\left[\frac{\alpha_{s}(\mu)}{\alpha_{s}({\rm 2GeV})}\right]^{\frac{12}{25}} \, ,\nonumber\\
\alpha_s(\mu)&=&\frac{1}{b_0t}\left[1-\frac{b_1}{b_0^2}\frac{\log t}{t} +\frac{b_1^2(\log^2{t}-\log{t}-1)+b_0b_2}{b_0^4t^2}\right]\, ,
\end{eqnarray}
   where $t=\log \frac{\mu^2}{\Lambda^2}$, $b_0=\frac{33-2n_f}{12\pi}$, $b_1=\frac{153-19n_f}{24\pi^2}$,
   $b_2=\frac{2857-\frac{5033}{9}n_f+\frac{325}{27}n_f^2}{128\pi^3}$,
   $\Lambda=210\,\rm{MeV}$, $292\,\rm{MeV}$  and  $332\,\rm{MeV}$ for the flavors
   $n_f=5$, $4$ and $3$, respectively  \cite{PDG,Narison-mix,Narison-Book}.

In this article, we choose three sets parameters,\\
{\bf A}. We evolve  the input parameters to the  energy scale  $ \mu =\sqrt{M_{P}^2-{\mathbb{M}}_c^2}$ to extract the masses $M_{P}$;\\
{\bf B}. We evolve  the input parameters to the  energy scale  $ \mu =1\,\rm{GeV}$ to extract the masses $M_{P}$;\\
{\bf C}. We evolve the input parameters except for $m_c(m_c)$ to the  energy scale  $ \mu =1\,\rm{GeV}$ to extract the masses $M_{P}$,
which is denoted by $ \mu =\overline{1.0}\,\rm{GeV}$. \\
For example, the parameters {\bf A} are chosen in Refs.\cite{WangPc,WangTetraquark}, the parameters {\bf B} are chosen in Refs.\cite{Wang-1GeV,Wang-1GeV-4430},
the parameters {\bf C} are chosen in Refs.\cite{Penta-anti-c,Penta-Nielsen,Nielsen-mc-1GeV}.

Now we take a short digression to discuss the energy scale formula, $ \mu =\sqrt{M_{P}^2-{\mathbb{M}}_Q^2}$. In the heavy quark limit,
the $Q$-quark serves as a static well potential and  can combine with a  light quark $q$  to form a heavy diquark  in  color antitriplet.
 The $\overline{Q}$-quark serves  as another static well potential, and can combine with a light diquark $\varepsilon^{ijk}q^i\,q^{\prime j}$ to form a heavy triquark in color triplet,
\begin{eqnarray}
q^j+Q^k &\to & \varepsilon^{ijk}\, q^j\,Q^k\, , \nonumber\\
\varepsilon^{ijl}q^i\,q^{\prime j} +\overline{Q}^k&\to & \varepsilon^{lkm}\varepsilon^{ijl}\,q^i \,q^{\prime j}\, \overline{Q}^k\, ,
\end{eqnarray}
where the $i$, $j$, $k$, $l$, $m$ are color indexes.
 Then
\begin{eqnarray}
 \varepsilon^{ijk}\, q^j\,Q^k+\varepsilon^{imn} \bar{q}^{\prime m}\,\overline{Q}^n &\to &  {\rm compact \,\,\, tetraquark \,\,\, states}\, , \nonumber\\
  \varepsilon^{lkm}\varepsilon^{ijl}\,q^i \,q^{\prime j}\, \overline{Q}^k+\varepsilon^{mnb} q^{\prime \prime n}\,Q^b &\to &
  {\rm compact \,\,\, pentaquark \,\,\, states}\, .
\end{eqnarray}
The tetraquark states $q\bar{q}^\prime Q \overline{Q}$ ($X,\,Y,\,Z$) and pentaquark states $qq^\prime q^{\prime\prime} Q \overline{Q}$ ($P$)
are characterized by the effective heavy quark masses ${\mathbb{M}}_Q$  and the virtuality
$V=\sqrt{M^2_{X/Y/Z}-(2{\mathbb{M}}_Q)^2}$, $\sqrt{M^2_{P}-(2{\mathbb{M}}_Q)^2}$. It is natural to take the energy  scales of the QCD spectral densities to be $\mu=V$.

The effective $Q$-quark masses ${\mathbb{M}}_Q$ have  universal values, and embody  the net effects of the complex dynamics
\cite{WangPc,WangTetraquark}. We fit  the   effective $Q$-quark masses ${\mathbb{M}}_{Q}$ to reproduce the experimental
values $M_{Z_c(3900)}$ and $M_{Z_b(10610)}$ in the  scenario of  tetraquark  states \cite{WangTetraquark},
 then use the  energy scale formula $\mu=\sqrt{M^2_{X/Y/Z}-(2{\mathbb{M}}_Q)^2}$ and
 $\sqrt{M^2_{P}-(2{\mathbb{M}}_c)^2}$ to study the
 hidden-charm (hidden-bottom) tetraquark states and  hidden-charm pentaquark states, respectively, and obtain satisfactory results.
    In this article,  we use the  formula $ \mu =\sqrt{M_{P}^2-{\mathbb{M}}_c^2}$ to determine the ideal energy scales of the QCD spectral densities, as
   there exists only one heavy quark, and  take the updated value of the effective $c$-quark mass ${\mathbb{M}}_c=1.82\,\rm{GeV}$ \cite{WangEPJC4260}.

We search for the ideal Borel parameters $T^2$ and continuum threshold parameters $s_0$   according to  the  four criteria:

$\bf{1_\cdot}$ Pole dominance at the hadron side;

$\bf{2_\cdot}$ Convergence of the operator product expansion;

$\bf{3_\cdot}$ Appearance of the Borel platforms;

$\bf{4_\cdot}$ Satisfying the energy scale formula $ \mu =\sqrt{M_{P}^2-{\mathbb{M}}_c^2}$ for the parameters {\bf A},  \\
by try and error,  and present the optimal energy scales $\mu$,  ideal Borel parameters $T^2$, continuum threshold parameters $s_0$ and
pole contributions  in Table 1. From Table 1, we can see that the criterion  $\bf{1}$ can be satisfied.

 In Fig.1, we plot the absolute  contributions of the vacuum condensates $|D(n)|$ in the operator product expansion  for the central values of the parameters shown
 in Table 1  in the case of the  parameters {\bf A},
   \begin{eqnarray}
D(n)&=& \frac{  \int_{m_c^2}^{s_0} ds\,\rho_{n}(s)\,\exp\left(-\frac{s}{T^2}\right)}{\int_{m_c^2}^{s_0} ds \,\rho(s)\,\exp\left(-\frac{s}{T^2}\right)}\, ,
\end{eqnarray}
where the $\rho_{n}(s)$ are the QCD spectral densities for the vacuum condensates of dimension $n$, and the total spectral densities
$\rho(s)=\sqrt{s}\rho^1_{QCD}(s)\pm \rho^0_{QCD}(s)$.
    From the figure, we can see that the dominant contributions come from the perturbative terms $D(0)$ for the positive parity pentaquark states,
    the operator product expansion is well convergent, while for the negative parity  pentaquark states, the contributions of the vacuum condensates of dimensions $n=10$, $12$,
    $13$ are tiny, the largest contributions come from the vacuum condensates of dimension $n=6$, but the contributions of the vacuum condensates of dimensions
    $6$, $8$, $9$, $11$ have the hierarchy $D(6)\gg |D(8)|\sim D(9)\gg |D(11)|$ or $D(6)\gg |D(8)|\gg D(9)\gg |D(11)|$, the operator product expansion is also convergent.
   On the other hand, from the figure, we can see that the  contributions of the perturbative terms $D_0$ are tiny for the negative parity pentaquark states,
   so in this article we approximate the continuum contributions as $\rho(s)\Theta(s-s_0)$, and define
    the pole contributions $\rm{PC}$ as
   \begin{eqnarray}
{\rm PC}&=& \frac{  \int_{m_c^2}^{s_0} ds\,\left[\sqrt{s}\rho_{QCD}^1(s)\pm  \rho_{QCD}^0(s)\right]\,
\exp\left(-\frac{s}{T^2}\right)}{\int_{m_c^2}^{\infty} ds \,\left[\sqrt{s}\rho_{QCD}^1(s)\pm \rho_{QCD}^0(s)\right]\,\exp\left(-\frac{s}{T^2}\right)}\, .
\end{eqnarray}
       In calculations, we observe that the dominant contributions come from the perturbative terms  $D(0)$ for the parameters shown in Table 1
       in the case of the parameters {\bf B} and {\bf C}, the operator product expansion are well convergent.  Now the criterion ${\bf 1}$ and criterion ${\bf 2}$ are satisfied, we expect to make reasonable predictions.

\begin{figure}
 \centering
 \includegraphics[totalheight=5cm,width=7cm]{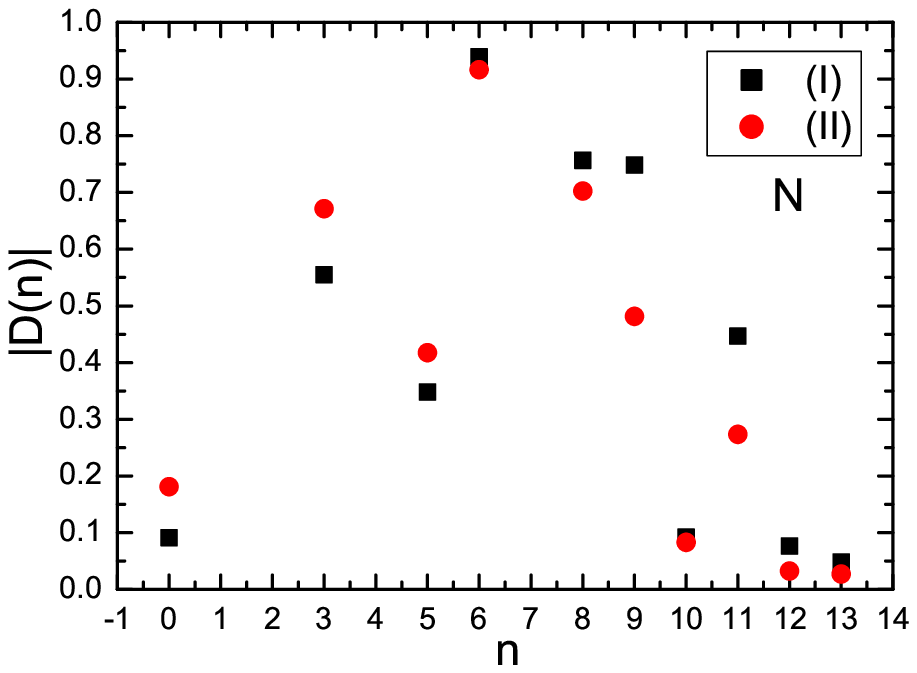}
  \includegraphics[totalheight=5cm,width=7cm]{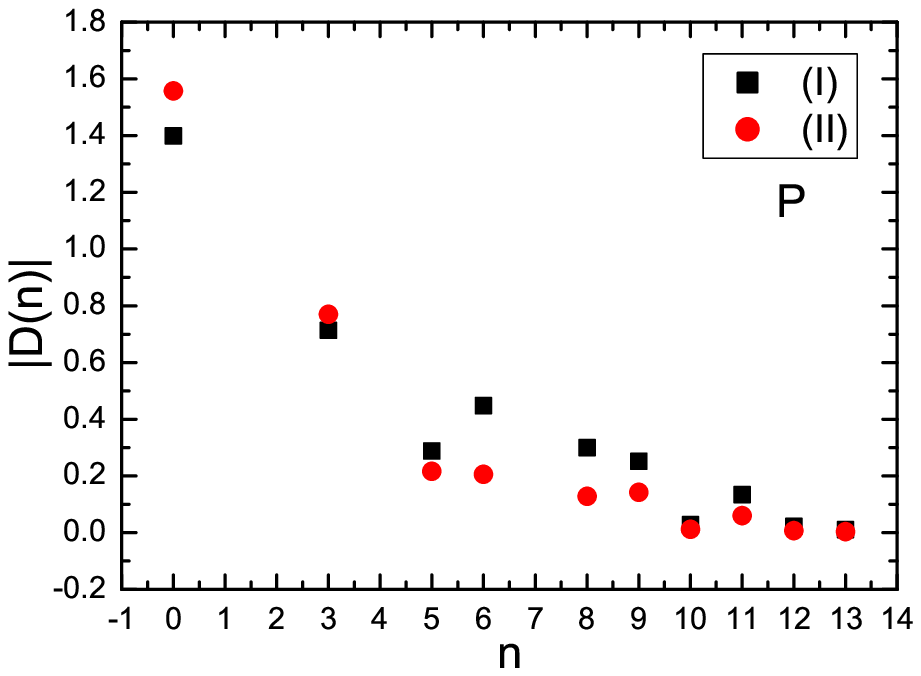}
            \caption{  The absolute contributions  of the vacuum condensates of dimension $n$ for central values of the  parameters {\bf A}, where  the (I) and
             (II)  denote the pentaquark states  $uuuc\bar{u}$ and $sssc\bar{s}$, respectively, the N and P denote
            the  negative parity and positive parity pentaquark states, respectively.   }
\end{figure}

We take into account all uncertainties  of the input    parameters,
and obtain  the masses and pole residues of
 the  charmed pentaquark  states with $J^P={\frac{3}{2}}^\pm$, which are shown explicitly in
Table 2. From Table 2, we can see that the criterion $\bf{4}$ can be satisfied for the parameters {\bf A}.  In Figs.2-7,
we plot the masses and pole residues of the charmed pentaquark states
 with variations
of the Borel parameters $T^2$ at much larger intervals   than the  Borel windows shown in Table 1. In the Borel windows, the uncertainties of the masses and pole residues
 originate from the Borel parameters $T^2$ are very small, the Borel platforms exist, the criterion $\bf{3}$ can be satisfied. Now the  four criteria are all satisfied,
and we expect to make
reliable predictions.

In the Borel windows, the uncertainties of the predicted masses are  less than $5\%$, as we obtain the masses from a ratio, see Eqs.(24-25), the
uncertainties originate from a special parameter in the  numerator and denominator    cancel   out   with each other,  the net uncertainties are very small;
 while the uncertainties of the pole residues can be as large as  $20\%$, as analogous cancelations do not exist.

If we choose  analogous  pole contributions, about $(40-60)\%$, the predicted masses based on the three sets  parameters
 have the relation, $M_{\bf A}<M_{\bf C}<M_{\bf B}$.   From  Table 2, we can see that for the negative parity charmed pentaquark states,
  the parameters {\bf A} lead to much smaller predicted masses than the parameters {\bf B} and {\bf C}.

\begin{table}
\begin{center}
\begin{tabular}{|c|c|c|c|c|c|c|c|}\hline\hline
                &$J^P$                    &$\mu(\rm GeV)$    &$T^2 (\rm{GeV}^2)$  &$\sqrt{s_0} (\rm{GeV})$   &pole                 \\  \hline
$uuuc\bar{u}$   &${\frac{3}{2}}^-$        &2.5               &$2.0-2.4$           &$3.6\pm0.1$               &$(38-65)\%$  \\
                &                         &1.0               &$2.7-3.1$           &$4.5\pm0.1$               &$(39-62)\%$          \\
                &                         &$\overline{1.0}$  &$2.5-2.9$           &$4.3\pm0.1$               &$(39-63)\%$          \\ \hline

$sssc\bar{s}$   &${\frac{3}{2}}^-$        &2.6               &$2.0-2.4$           &$3.7\pm0.1$               &$(38-65)\%$  \\
                &                         &1.0               &$2.8-3.2$           &$4.6\pm0.1$               &$(39-61)\%$          \\
                &                         &$\overline{1.0}$  &$2.7-3.1$           &$4.5\pm0.1$               &$(40-63)\%$          \\ \hline

$uuuc\bar{u}$   &${\frac{3}{2}}^+$        &4.1               &$3.2-3.6$           &$5.0\pm0.1$               &$(39-60)\%$  \\
                &                         &1.0               &$3.1-3.5$           &$5.1\pm0.1$               &$(38-60)\%$          \\
                &                         &$\overline{1.0}$  &$3.2-3.6$           &$5.1\pm0.1$               &$(39-60)\%$          \\ \hline

$sssc\bar{s}$   &${\frac{3}{2}}^+$        &4.3               &$3.4-3.8$           &$5.2\pm0.1$               &$(40-59)\%$  \\
                &                         &1.0               &$3.2-3.6$           &$5.3\pm0.1$               &$(40-61)\%$          \\
                &                         &$\overline{1.0}$  &$3.3-3.7$           &$5.3\pm0.1$               &$(41-62)\%$          \\ \hline  \hline
\end{tabular}
\end{center}
\caption{ The optimal energy scales $\mu$, Borel parameters $T^2$, continuum threshold parameters $s_0$ and
 pole contributions (pole)    for the charmed pentaquark states.}
\end{table}

\begin{table}
\begin{center}
\begin{tabular}{|c|c|c|c|c|c|c|c|}\hline\hline
                &$J^P$                    &$\mu(\rm GeV)$    &$M (\rm{GeV})$          &$\lambda (\rm{GeV}^6)$             \\  \hline
$uuuc\bar{u}$   &${\frac{3}{2}}^-$        &2.5               &$3.07^{+0.13}_{-0.14}$  &$6.02^{+1.22}_{-1.04} \times 10^{-4}$      \\
                &                         &1.0               &$3.97^{+0.11}_{-0.12}$  &$1.35^{+0.22}_{-0.20} \times 10^{-3}$        \\
                &                         &$\overline{1.0}$  &$3.80^{+0.10}_{-0.12}$  &$1.16^{+0.19}_{-0.16} \times 10^{-3}$       \\ \hline

$sssc\bar{s}$   &${\frac{3}{2}}^-$        &2.6               &$3.22^{+0.12}_{-0.14}$  &$6.88^{+1.35}_{-1.13} \times 10^{-4}$      \\
                &                         &1.0               &$4.07^{+0.10}_{-0.10}$  &$1.62^{+0.27}_{-0.23} \times 10^{-3}$        \\
                &                         &$\overline{1.0}$  &$3.96^{+0.10}_{-0.10}$  &$1.59^{+0.26}_{-0.23} \times 10^{-3}$       \\ \hline

$uuuc\bar{u}$   &${\frac{3}{2}}^+$        &4.1               &$4.49^{+0.13}_{-0.08}$  &$2.71^{+0.41}_{-0.32} \times 10^{-3}$      \\
                &                         &1.0               &$4.56^{+0.09}_{-0.08}$  &$1.97^{+0.33}_{-0.28} \times 10^{-3}$        \\
                &                         &$\overline{1.0}$  &$4.53^{+0.09}_{-0.09}$  &$2.23^{+0.36}_{-0.31} \times 10^{-3}$       \\ \hline

$sssc\bar{s}$   &${\frac{3}{2}}^+$        &4.3               &$4.70^{+0.16}_{-0.11}$  &$3.44^{+0.56}_{-0.42} \times 10^{-3}$      \\
                &                         &1.0               &$4.80^{+0.11}_{-0.11}$  &$2.54^{+0.38}_{-0.35} \times 10^{-3}$        \\
                &                         &$\overline{1.0}$  &$4.76^{+0.09}_{-0.10}$  &$2.83^{+0.42}_{-0.38} \times 10^{-3}$       \\ \hline  \hline
\end{tabular}
\end{center}
\caption{ The predicted masses and pole residues of the charmed pentaquark states.}
\end{table}

\begin{figure}
 \centering
 \includegraphics[totalheight=5cm,width=7cm]{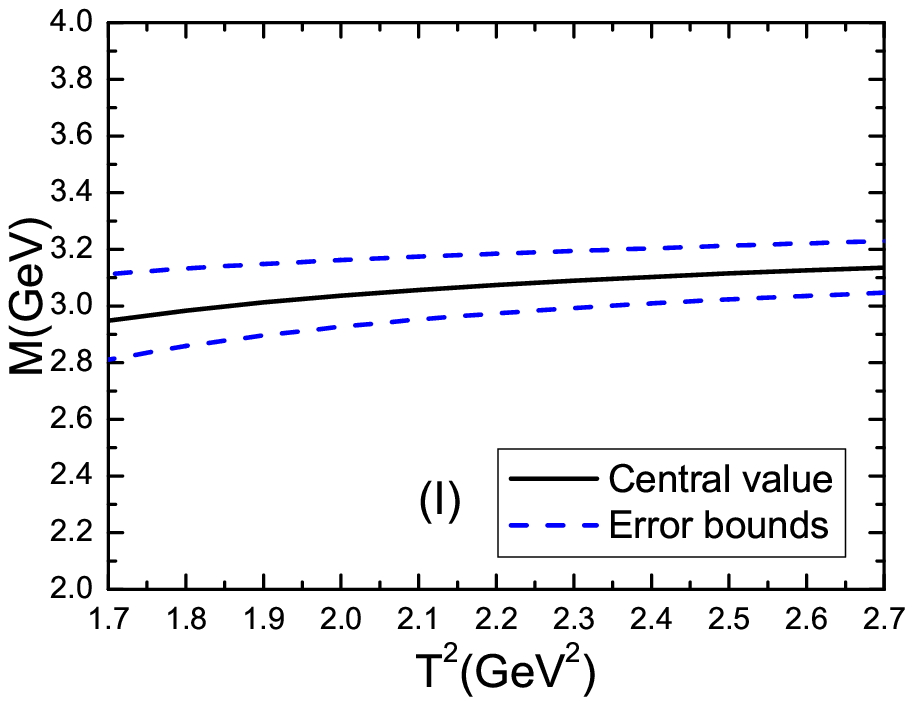}
  \includegraphics[totalheight=5cm,width=7cm]{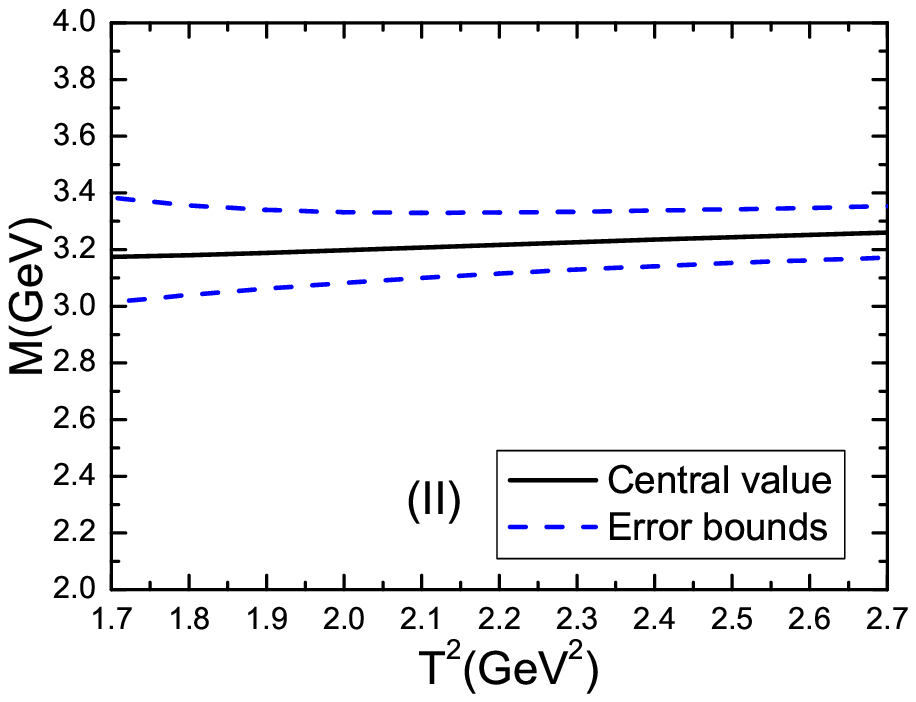}
   \includegraphics[totalheight=5cm,width=7cm]{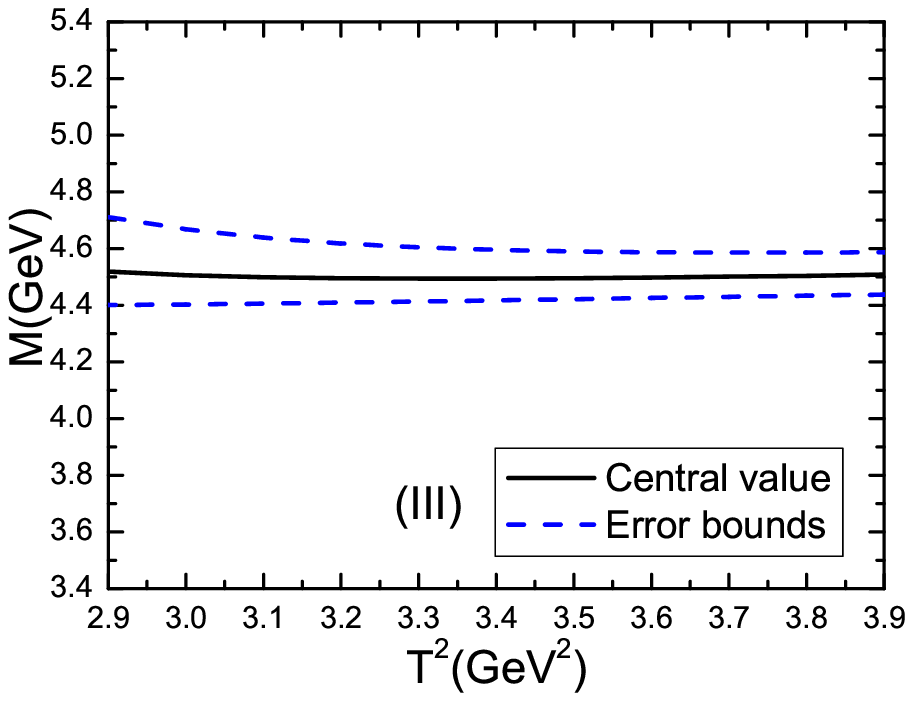}
    \includegraphics[totalheight=5cm,width=7cm]{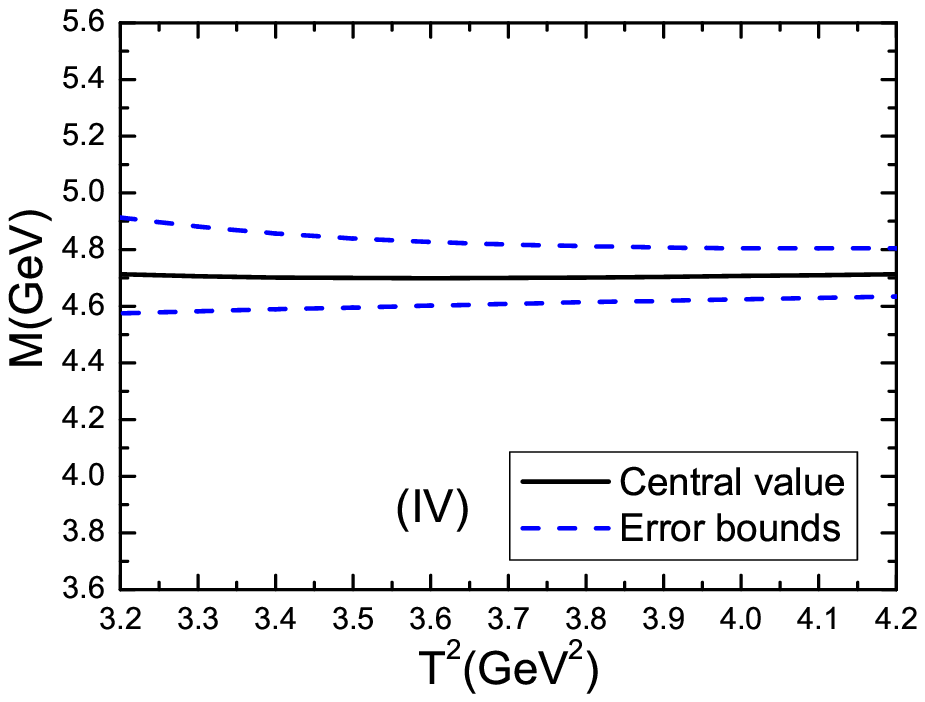}
         \caption{ The masses  of the charmed pentaquark states   with variations of the Borel parameters $T^2$ for the parameters {\bf A}, where the (I), (II), (III) and (IV)
                  correspond to the  quantum numbers $ (uuuc\bar{u},{\frac{3}{2}}^-)$, $ (sssc\bar{s},{\frac{3}{2}}^-)$,
         $ (uuuc\bar{u},{\frac{3}{2}}^+)$,
          and $ (sssc\bar{s},{\frac{3}{2}}^+)$, respectively.   }
\end{figure}

\begin{figure}
 \centering
 \includegraphics[totalheight=5cm,width=7cm]{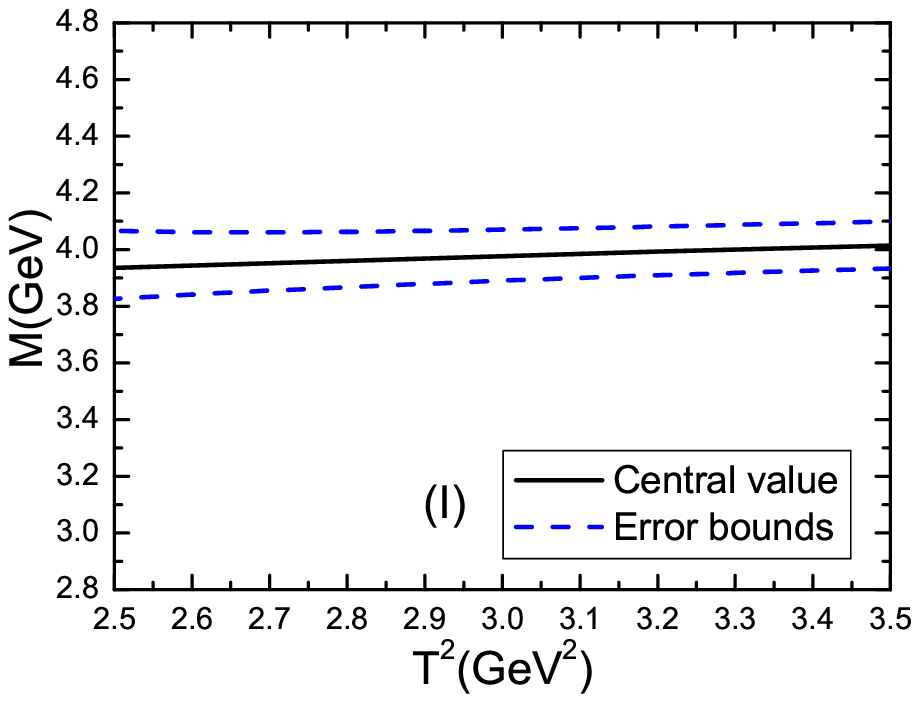}
  \includegraphics[totalheight=5cm,width=7cm]{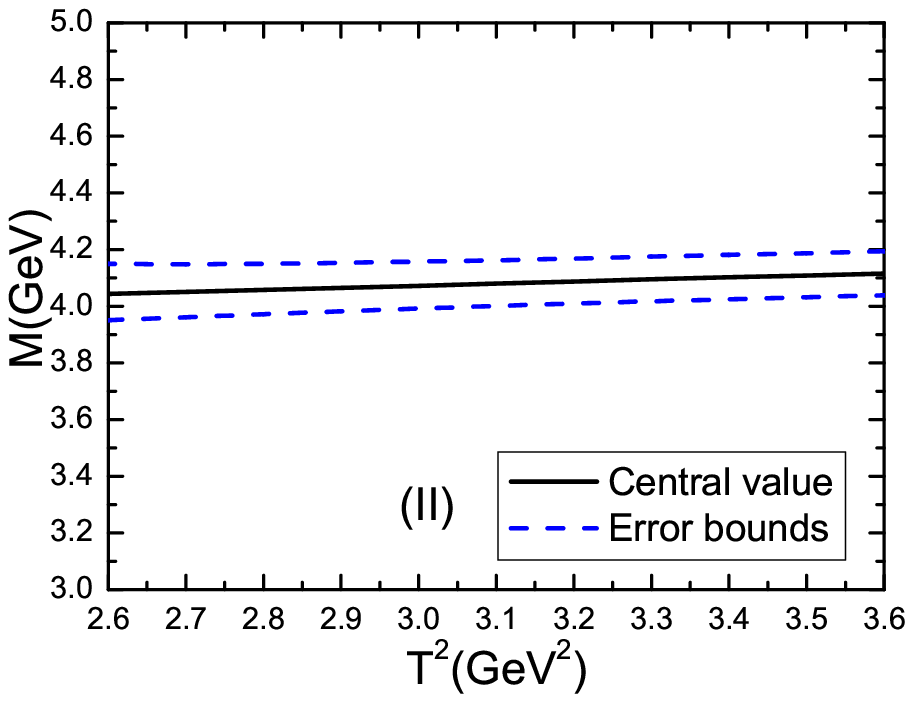}
   \includegraphics[totalheight=5cm,width=7cm]{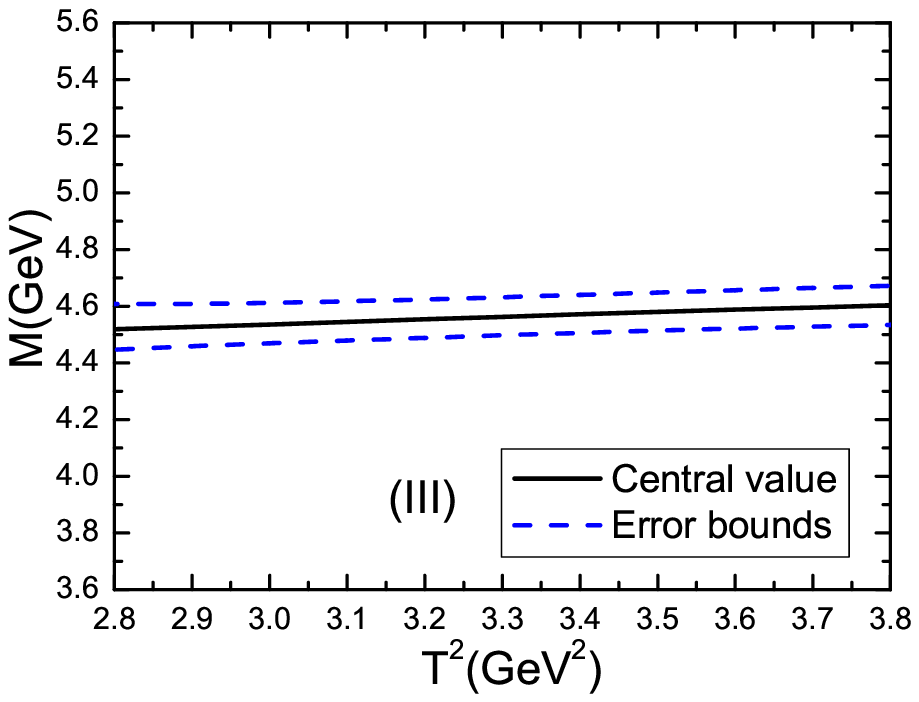}
    \includegraphics[totalheight=5cm,width=7cm]{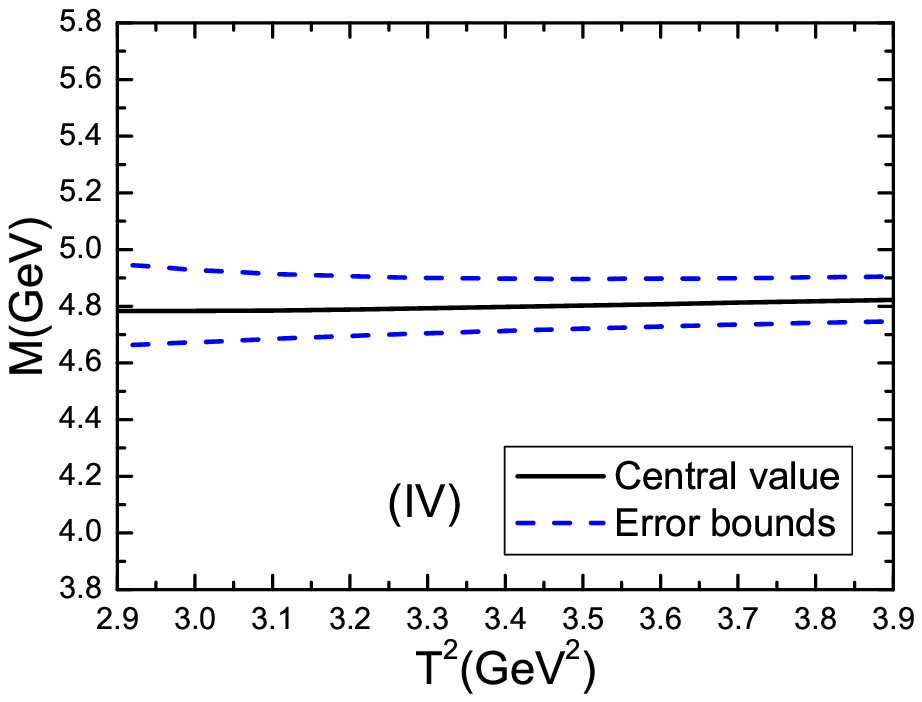}
         \caption{ The masses  of the charmed pentaquark states   with variations of the Borel parameters $T^2$ for the parameters {\bf B}, where the (I), (II), (III) and (IV)
                  correspond to the  quantum numbers $ (uuuc\bar{u},{\frac{3}{2}}^-)$, $ (sssc\bar{s},{\frac{3}{2}}^-)$,
         $ (uuuc\bar{u},{\frac{3}{2}}^+)$,
          and $ (sssc\bar{s},{\frac{3}{2}}^+)$, respectively.   }
\end{figure}

\begin{figure}
 \centering
 \includegraphics[totalheight=5cm,width=7cm]{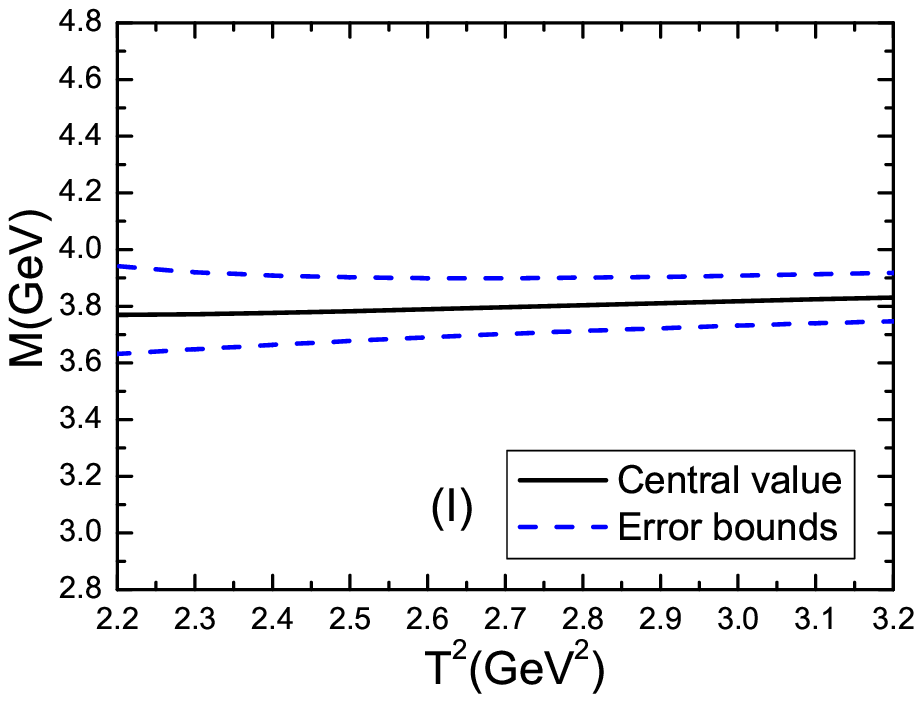}
  \includegraphics[totalheight=5cm,width=7cm]{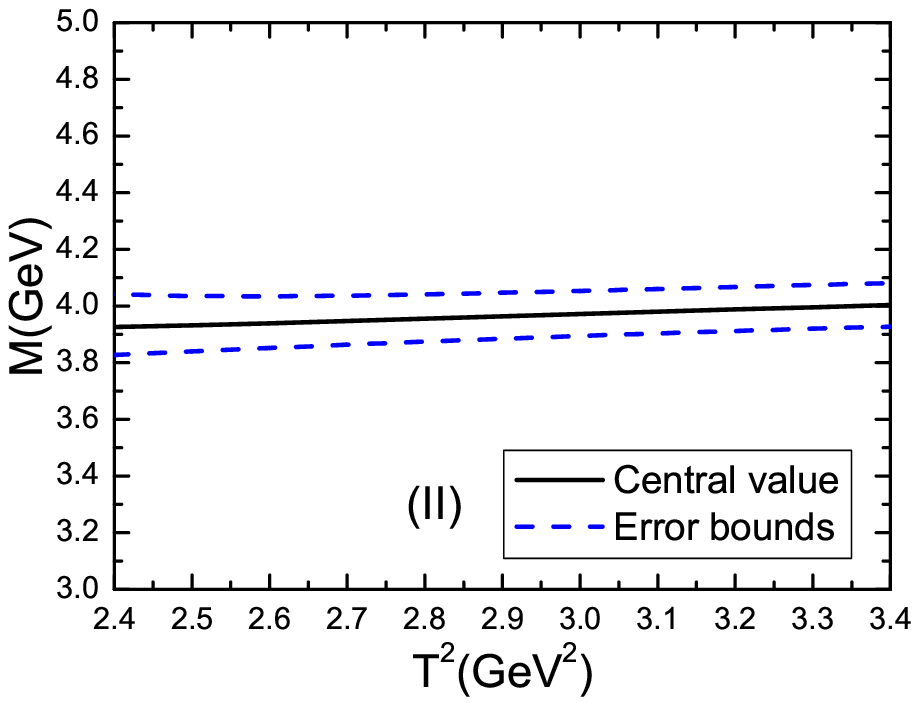}
   \includegraphics[totalheight=5cm,width=7cm]{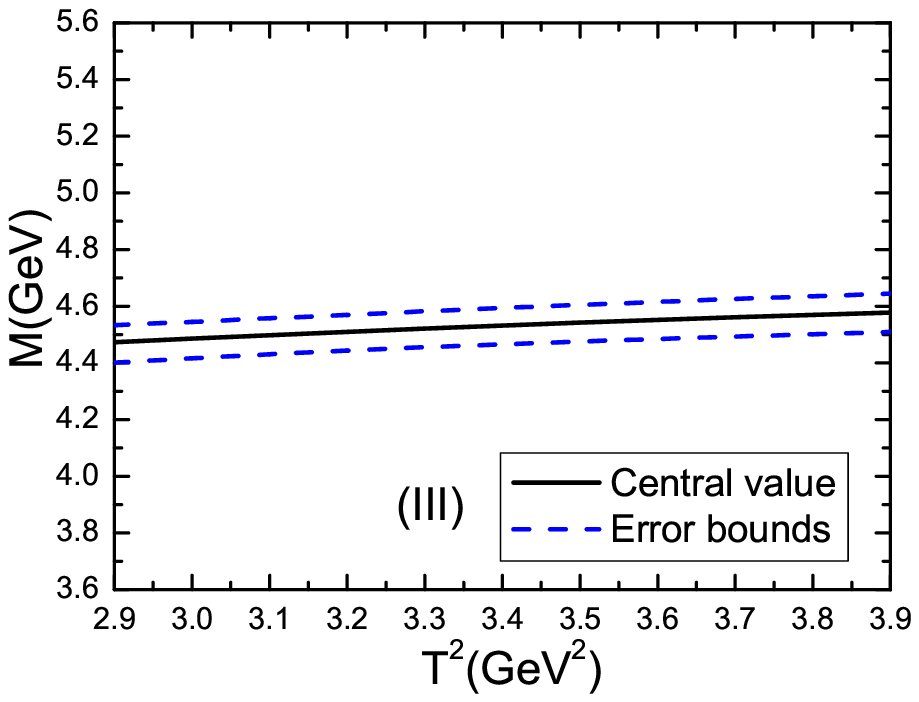}
    \includegraphics[totalheight=5cm,width=7cm]{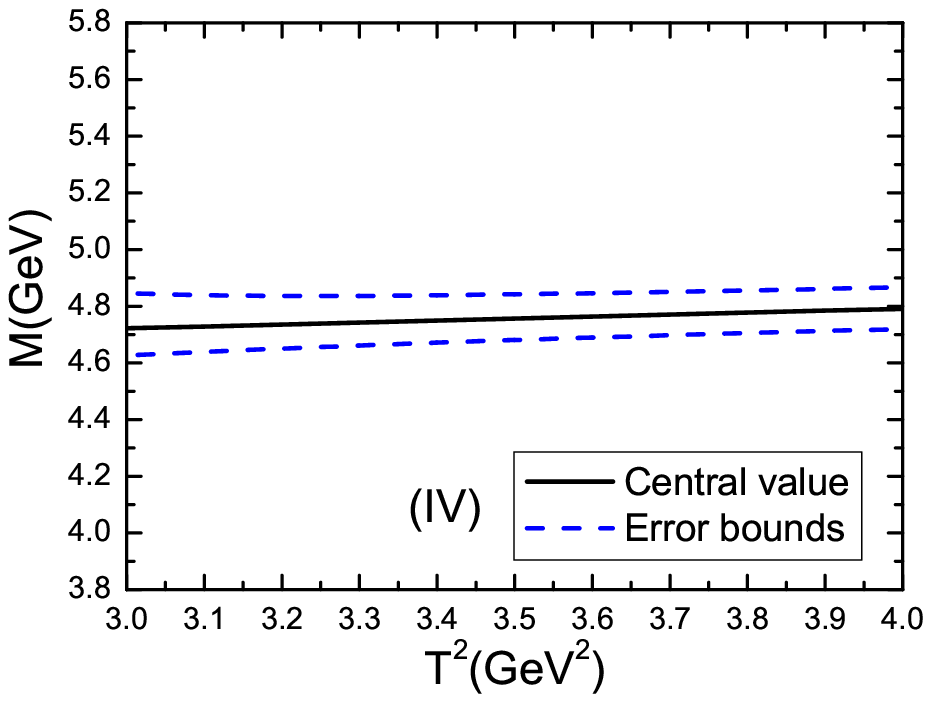}
         \caption{ The masses  of the charmed pentaquark states   with variations of the Borel parameters $T^2$ for the parameters {\bf C}, where the (I), (II), (III) and (IV)
                  correspond to the  quantum numbers $ (uuuc\bar{u},{\frac{3}{2}}^-)$, $ (sssc\bar{s},{\frac{3}{2}}^-)$,
         $ (uuuc\bar{u},{\frac{3}{2}}^+)$,
          and $ (sssc\bar{s},{\frac{3}{2}}^+)$, respectively.   }
\end{figure}

\begin{figure}
 \centering
 \includegraphics[totalheight=5cm,width=7cm]{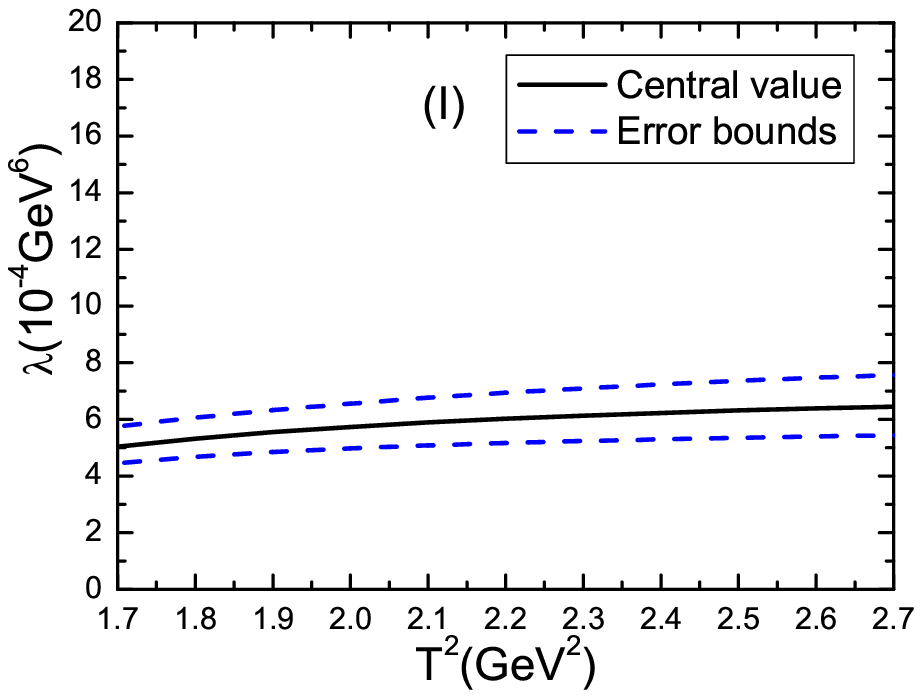}
  \includegraphics[totalheight=5cm,width=7cm]{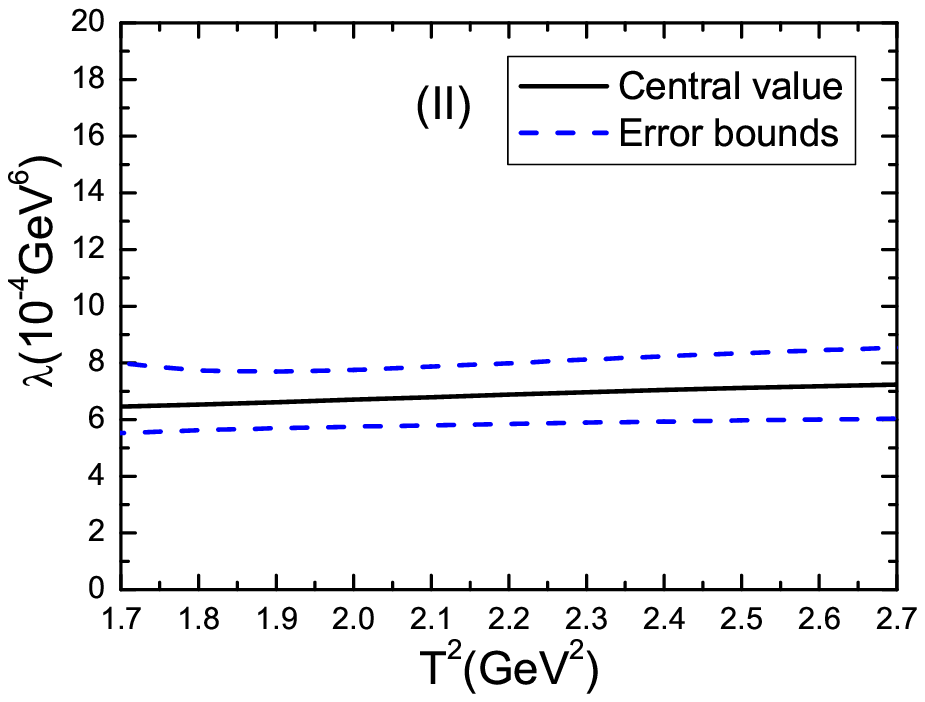}
   \includegraphics[totalheight=5cm,width=7cm]{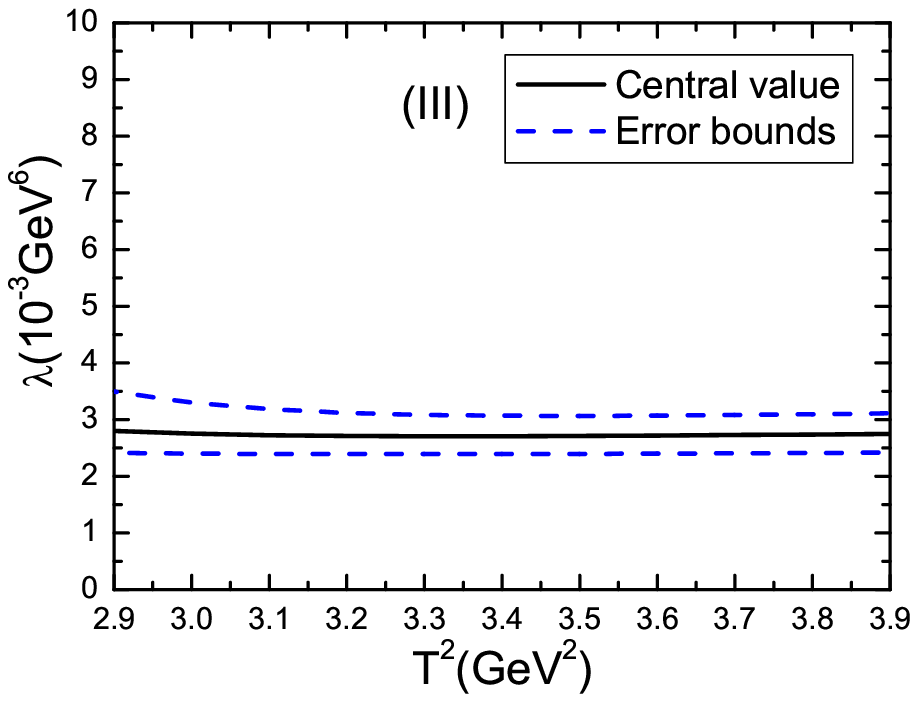}
    \includegraphics[totalheight=5cm,width=7cm]{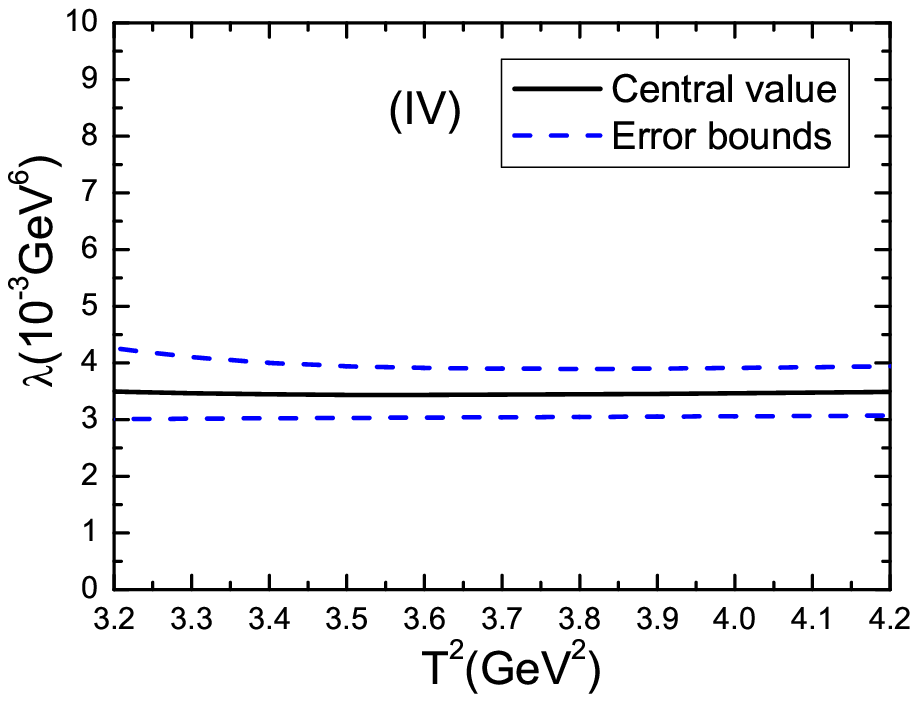}
         \caption{ The pole residues   of the charmed pentaquark states   with variations of the Borel parameters $T^2$ for the parameters {\bf A}, where the (I), (II), (III) and (IV)
                  correspond to the  quantum numbers $ (uuuc\bar{u},{\frac{3}{2}}^-)$, $ (sssc\bar{s},{\frac{3}{2}}^-)$,
         $ (uuuc\bar{u},{\frac{3}{2}}^+)$,
          and $ (sssc\bar{s},{\frac{3}{2}}^+)$, respectively.   }
\end{figure}

\begin{figure}
 \centering
 \includegraphics[totalheight=5cm,width=7cm]{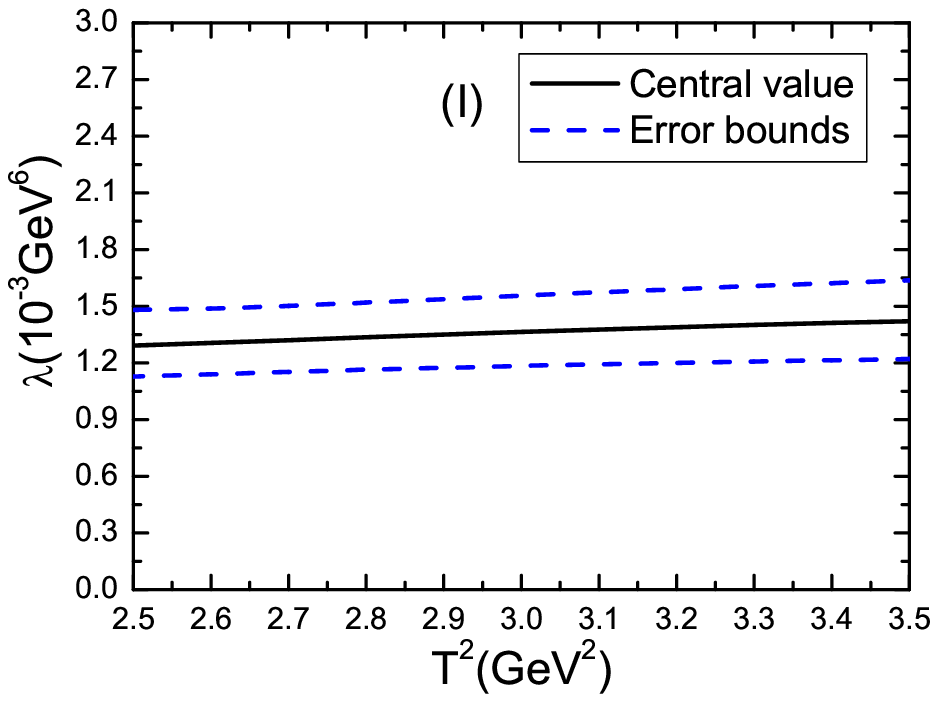}
  \includegraphics[totalheight=5cm,width=7cm]{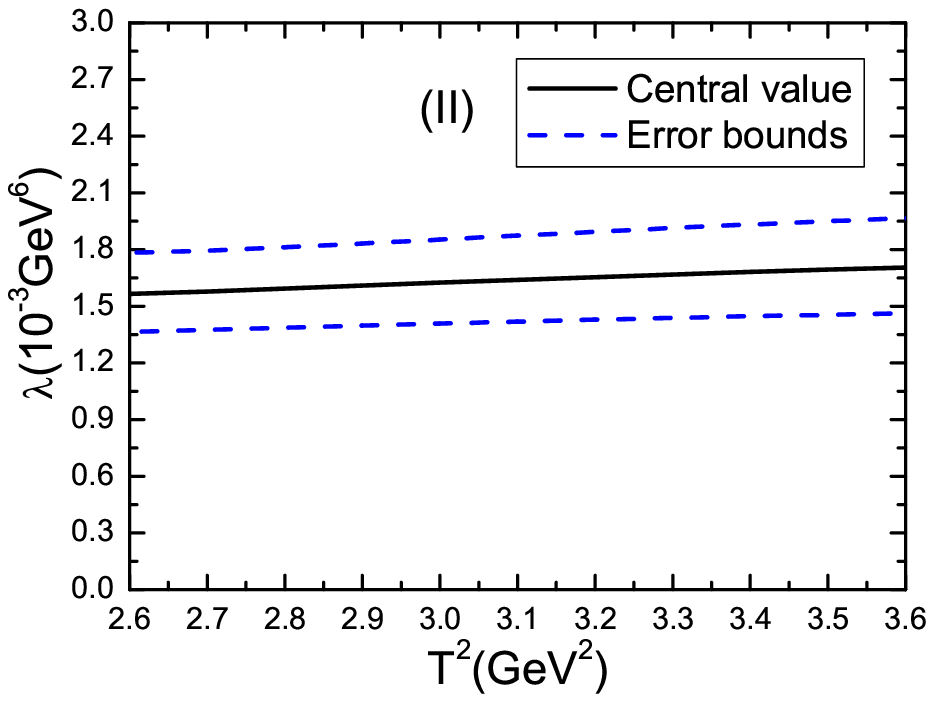}
   \includegraphics[totalheight=5cm,width=7cm]{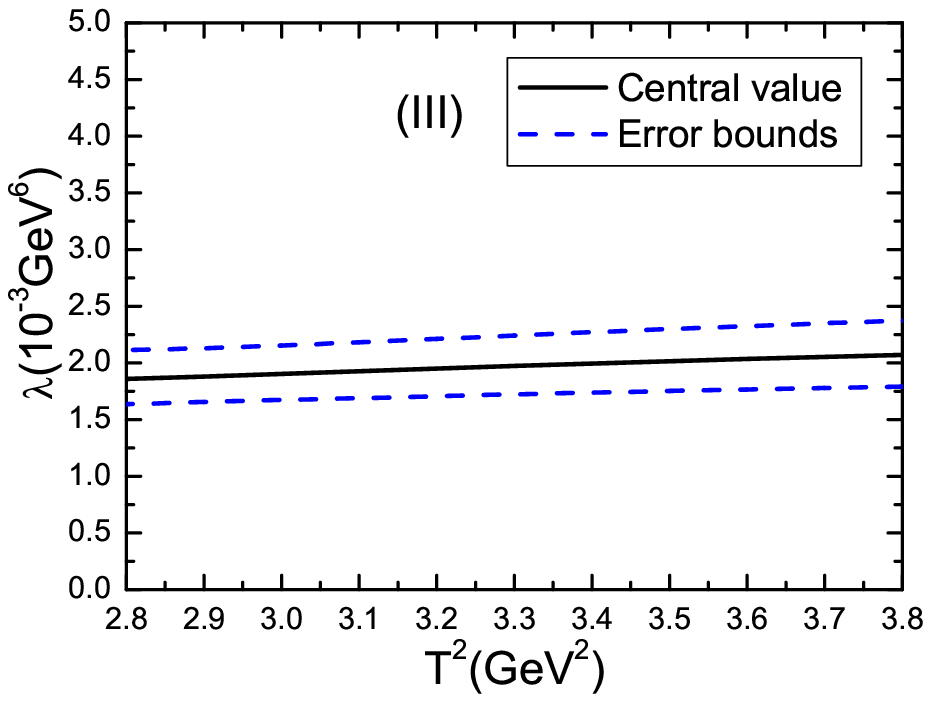}
    \includegraphics[totalheight=5cm,width=7cm]{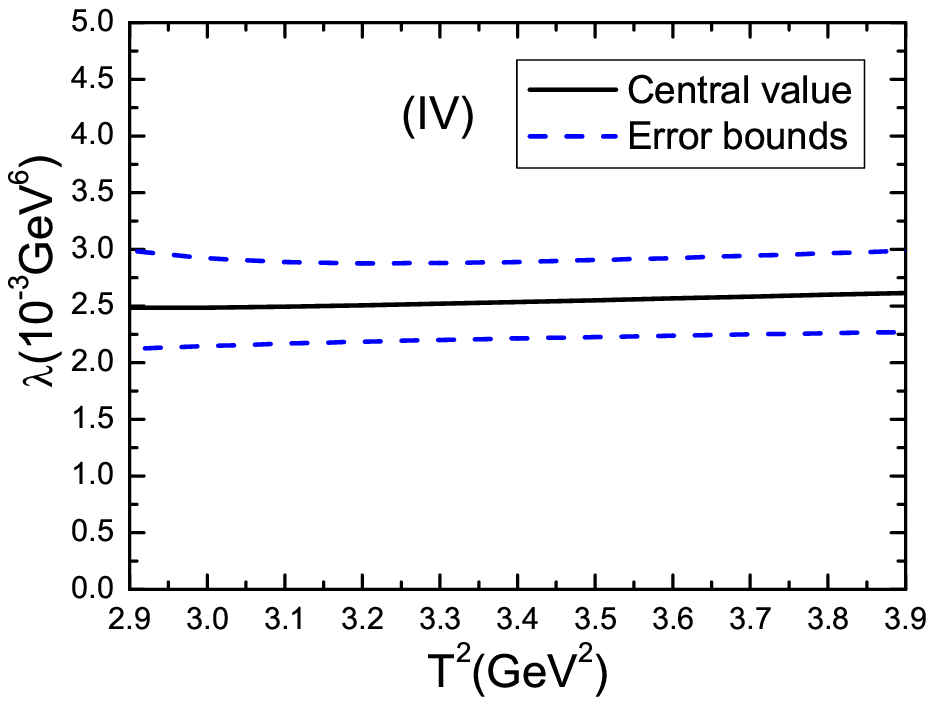}
         \caption{ The pole residues   of the charmed pentaquark states   with variations of the Borel parameters $T^2$ for the parameters {\bf B}, where the (I), (II), (III) and (IV)
                  correspond to the  quantum numbers $ (uuuc\bar{u},{\frac{3}{2}}^-)$, $ (sssc\bar{s},{\frac{3}{2}}^-)$,
         $ (uuuc\bar{u},{\frac{3}{2}}^+)$,
          and $ (sssc\bar{s},{\frac{3}{2}}^+)$, respectively.   }
\end{figure}

\begin{figure}
 \centering
 \includegraphics[totalheight=5cm,width=7cm]{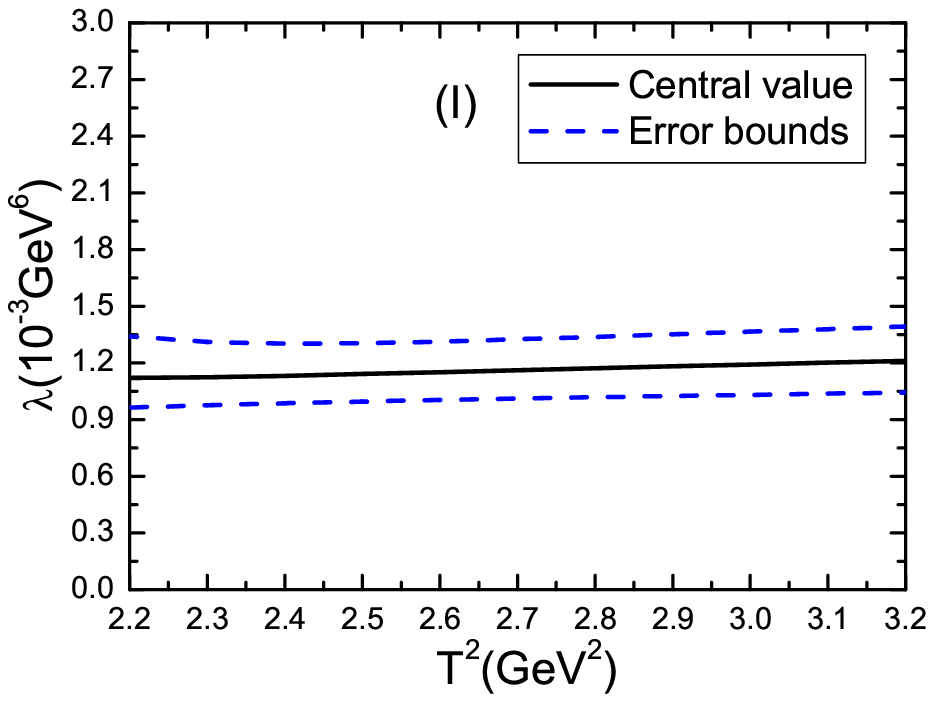}
  \includegraphics[totalheight=5cm,width=7cm]{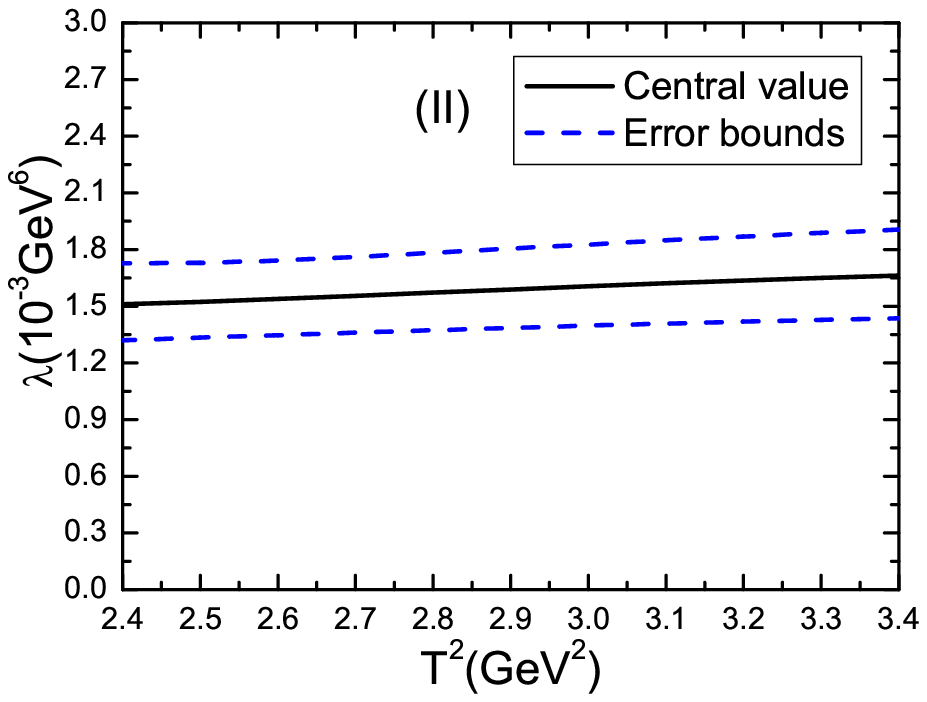}
   \includegraphics[totalheight=5cm,width=7cm]{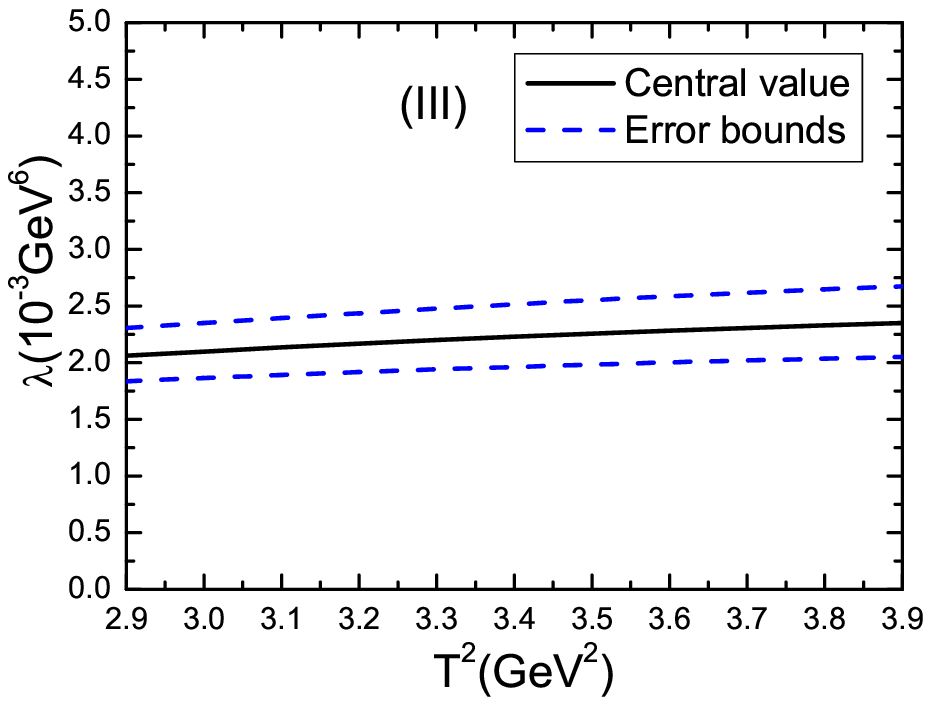}
    \includegraphics[totalheight=5cm,width=7cm]{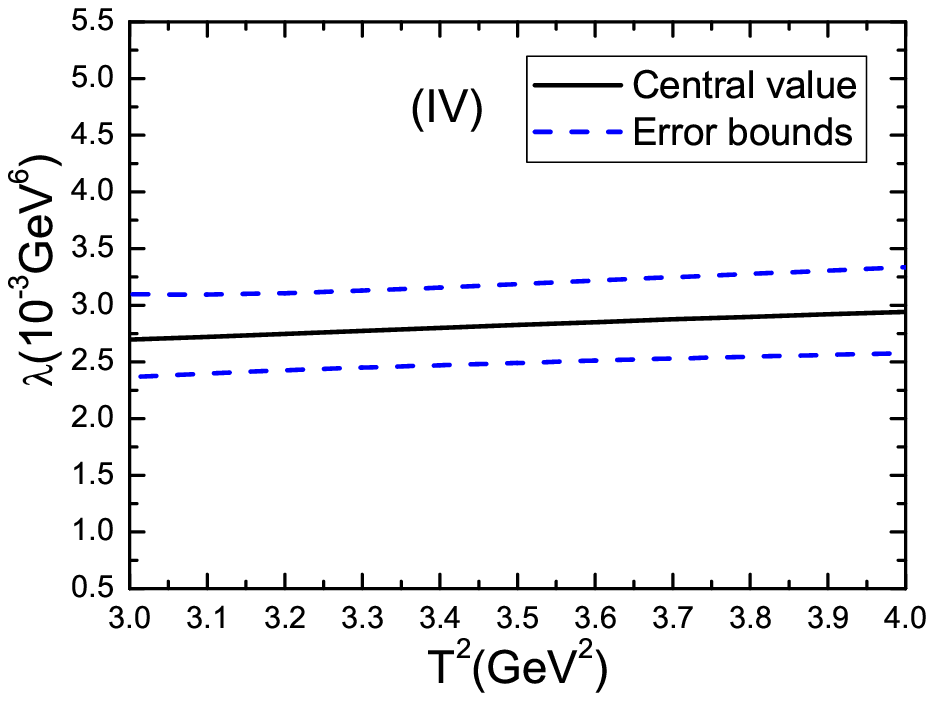}
         \caption{ The pole residues   of the charmed pentaquark states   with variations of the Borel parameters $T^2$ for the parameters {\bf C}, where the (I), (II), (III) and (IV)
                  correspond to the  quantum numbers $ (uuuc\bar{u},{\frac{3}{2}}^-)$, $ (sssc\bar{s},{\frac{3}{2}}^-)$,
         $ (uuuc\bar{u},{\frac{3}{2}}^+)$,
          and $ (sssc\bar{s},{\frac{3}{2}}^+)$, respectively.   }
\end{figure}

In Ref.\cite{Penta-Nielsen},  Albuquerque, Lee and Nielsen  study the charmed pentaquark states $udcd\bar{u}$ with $J^P={\frac{1}{2}}^+$ with the QCD sum rules  by
taking into account the vacuum condensates up to dimension $10$, and obtain the ground state masses $3.21 \pm 0.13\,\rm{GeV}$ and $4.15 \pm 0.11\,\rm{GeV}$ for the
scalar-diquark-scalar-diquark-antiquark type and scalar-diquark-pseudoscalar-diquark-antiquark type pentaquark states, respectively. In Ref.\cite{Penta-Nielsen}, the parameters
{\bf C} are chosen, if we choose the parameters
{\bf C}, we obtain the prediction $M=4.53\pm 0.09\,\rm{GeV}$ for the axialvector-diquark-scalar-diquark-antiquark type pentaquark state $uuuc\bar{u}$
with $J^P={\frac{3}{2}}^+$.  The calculations  based on the QCD sum rules indicate that the axialvector light diquark states $A$ have larger masses than the corresponding scalar light
diquark states $S$, $M_{A}-M_{S}=0.15\sim 0.20\,\rm{GeV}$  \cite{Wang-light-Diquark}.
We can estimate that the scalar-diquark-scalar-diquark-antiquark type pentaquark state  $udcd\bar{u}$ with $J^P={\frac{1}{2}}^+$ has a mass about $4.36\pm0.09\,\rm{GeV}$, which
is much larger than the value $3.21 \pm 0.13\,\rm{GeV}$ obtained in Ref.\cite{Penta-Nielsen} in a Borel window
 where the contributions of the vacuum condensates of dimension $10$ are still very large, the convergent behavior of the operator product expansion is very bad.   In this article, we carry out the operator product expansion up to the
 vacuum condensates of  dimension $13$ in a consistent way. We do not prefer  the parameters
{\bf C} as they lead to  two energy scales, $\mu=m_c$ and $\mu=1\,\rm{GeV}$,  in the QCD spectral densities.

 The predicated masses depend on the input parameters {\bf A}, {\bf B} and {\bf C}, see Table 2. In Ref.\cite{Wang-1GeV-4430},
 we obtain the mass $M_Z=4.44 \pm 0.19\,\rm{GeV}$ for the $Z(4430)$ as the ground state  diquark-antidiquark type axialvector tetraquark state based on the QCD sum rules
 for the parameters {\bf B}. While in Ref.\cite{Wang-Z3900-Z4430},  we observe that  the $Z_c(3900)$ and $Z(4430)$ can be tentatively assigned to be the ground
 state and the first radial excited state of the diquark-antidiquark  type axialvector tetraquark states respectively for the parameters {\bf A}.
 In Ref.\cite{WangZ3900-Decay},  we  assign the $Z_c(3900)$ to be  the diquark-antidiquark  type axialvector  tetraquark  state,
  study its width with the QCD sum rules by taking into account all the Feynman diagrams for  the parameters {\bf A}, and reproduce the experimental value.
    From Refs.\cite{Wang-1GeV-4430,Wang-Z3900-Z4430} and present work, we can see that the parameters {\bf A} lead to smaller or much smaller masses than the parameters {\bf B}.

 \begin{figure}
 \centering
 \includegraphics[totalheight=5cm,width=7cm]{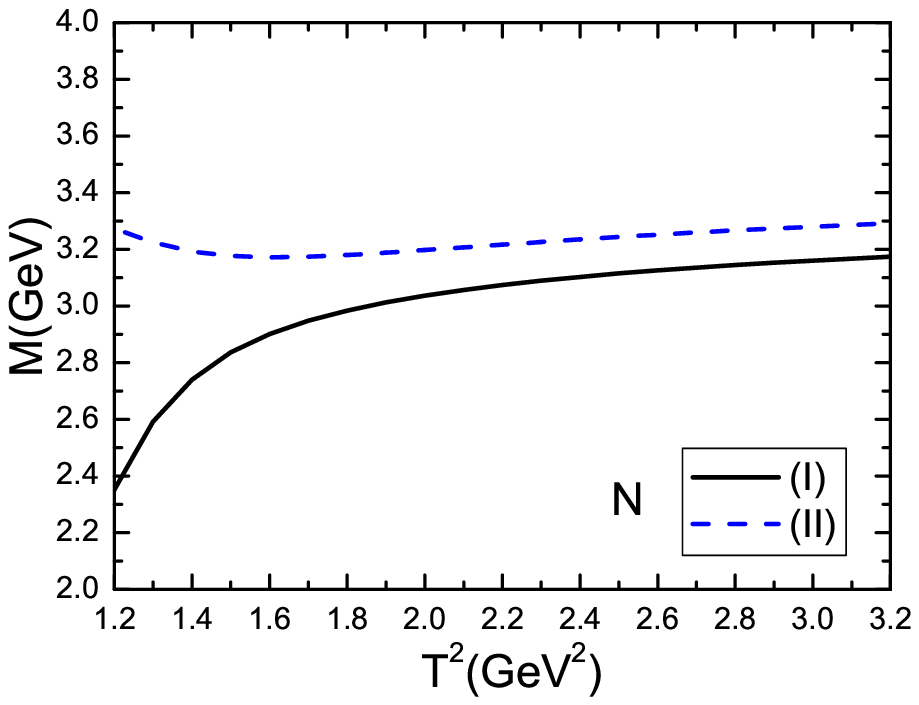}
  \includegraphics[totalheight=5cm,width=7cm]{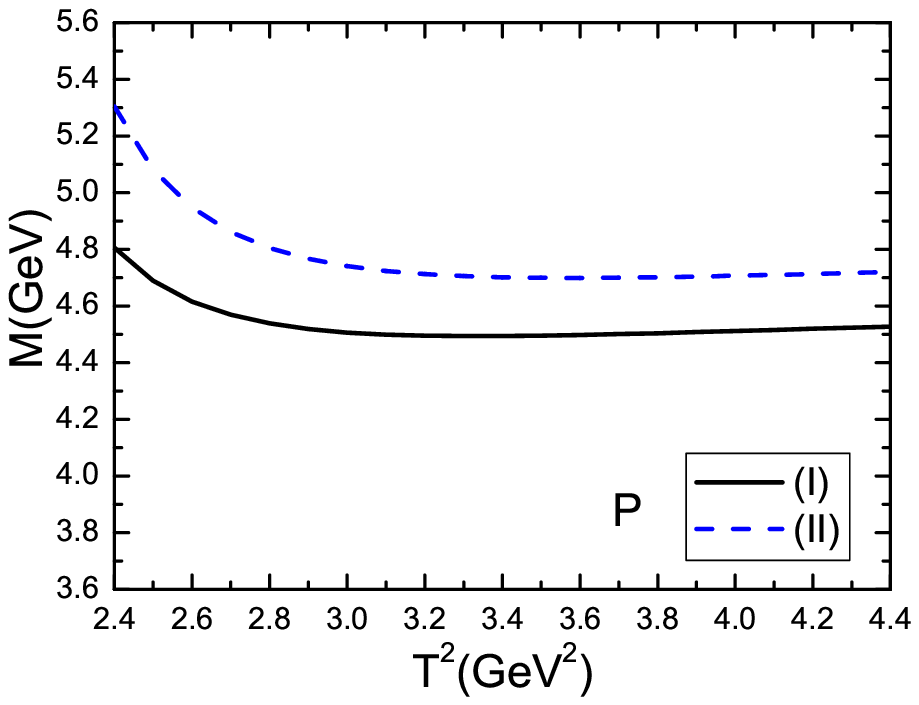}
            \caption{  The predicted masses with variations   of the Borel parameters for central values of the  parameters {\bf A}, where  the (I) and
             (II)  denote the pentaquark states  $uuuc\bar{u}$ and $sssc\bar{s}$, respectively, the N and P denote
            the  negative parity and positive parity pentaquark states, respectively.   }
\end{figure}

 In Fig.8, we plot the predicted masses with variations   of the Borel parameters for central values of the  parameters {\bf A} at very large intervals. From the figure,
 we can see that for small Borel parameters,  the predicted masses of the negative parity pentaquark state $sssc\bar{s}$   and positive parity pentaquark states
  $uuuc\bar{u}$ and $sssc\bar{s}$ increase  monotonously with the decrease of the Borel parameters, which warrant appearance  of very flat Borel platforms,
  as the predicted masses
  always increase monotonously with the increase of the Borel parameters for large Borel parameters.

From Table 2, we can see that the predicted masses of the pentaquark states $uuuc\bar{u}$ and $sssc\bar{s}$  with $J^P={\frac{3}{2}}^-$   are
$3.07^{+0.13}_{-0.14}\,\rm{GeV}$   and          $3.22^{+0.12}_{-0.14}\,\rm{GeV}$ respectively based on the parameters {\bf A}.
The masses of the pentaquark states  $ssuc\bar{u}$, $susc\bar{u}$, $ssdc\bar{d}$ and $sdsc\bar{d}$ can be estimated to be
\begin{eqnarray}
M&=&\frac{M_{uuuc\bar{u}}+M_{sssc\bar{s}}}{2}=3.15\pm0.13\,\rm{GeV}\, ,
\end{eqnarray}
which lie in the same region of the masses of the $\Omega_c(3050)$, $\Omega_c(3066)$, $\Omega_c(3090)$, $\Omega_c(3119)$ from the LHCb collaboration \cite{LHCb-Omega}.
In Ref.\cite{Pneta-Anisovich}, Anisovich et al obtain the mass  $M=3.2\pm0.1\,\rm{GeV}$ for the pentaquark state $ussc\bar{u}$ with $J^P={\frac{3}{2}}^-$ based on
the diquark-diquark-antiquark model, which is consistent with the present predictions. The new excited $\Omega_c$ states are possible candidates for the charmed  pentaquark
states, more experimental and theoretical works are still needed to make a solid assignment.
In this article, we prefer the parameters {\bf A},  because the  parameters {\bf A} can enhance the pole contributions remarkably
and improve the convergent behaviors significantly in the operator
product expansion  in the QCD sum rules for the exotic hadrons, such as the tetraquark states, pentaquark states, molecular states,
and  lead to much smaller predicted masses than the parameters {\bf B} and {\bf C}. We can assign more exotic hadrons reasonably based on the QCD sum rules if the
 parameters {\bf A} are chosen. However,  the parameters {\bf B} and {\bf C} are not excluded, more experimental data are still needed to select   the best parameters.

\section{Conclusion}

In this article, we  focus on  the  scenario of pentaquark  states interpretation of the new excited $\Omega_c$ states, and study
 the $J^P={\frac{3}{2}}^\pm$ charmed pentaquark states
with the QCD sum rules by carrying out the operator product expansion   up to   the vacuum condensates of dimension $13$ in a consistent way.
In calculations,   we separate  the contributions of the negative parity and positive parity pentaquark states  unambiguously,
and choose three sets input parameters to study the masses and pole residues of the charmed pentaquark states $uuuc\bar{u}$ and $sssc\bar{s}$
 with the QCD sum rules in details.  Then we estimate the masses of the charmed pentaquark states $ssuc\bar{u}$, $susc\bar{u}$, $ssdc\bar{d}$ and $sdsc\bar{d}$\
  with $J^P={\frac{3}{2}}^-$ to be $3.15\pm0.13\,\rm{GeV}$ according to the $SU(3)$ breaking effects, which is compatible with the experimental values of
  the masses of the $\Omega_c(3050)$, $\Omega_c(3066)$, $\Omega_c(3090)$,
  $\Omega_c(3119)$. The new excited $\Omega_c$ states are possible candidates for the charmed  pentaquark
states, more experimental and theoretical works are still needed to make a solid assignment.

\section*{Acknowledgements}
This  work is supported by National Natural Science Foundation, Grant Number 11775079.

\end{document}